\documentclass[12pt]{article}
\usepackage{bm}
\usepackage{natbib}
\usepackage{xcolor}
\usepackage{amsfonts}
\usepackage{amsmath}
\usepackage{amssymb}
\usepackage{mathrsfs}
\usepackage{hyperref}
\usepackage{booktabs}
\usepackage{graphicx}
\usepackage{multirow, multicol}
\usepackage[ruled,vlined,lined,boxed,linesnumbered]{algorithm2e}
\usepackage{enumitem}

\newtheorem{lemma}{\sc Lemma}
\newtheorem{theorem}{\sc Theorem}
\newtheorem{corollary}{\sc Corollary}
\newtheorem{assumption}{\sc Assumption}

\newtheorem{prop}{\sc Proposition}

\newcommand{\bit}{\begin{itemize}}
\newcommand{\eit}{\end{itemize}}

\newcommand{\ba}{\begin{eqnarray}}
\newcommand{\ea}{\end{eqnarray}}
\newcommand{\bas}{\begin{eqnarray*}}
\newcommand{\eas}{\end{eqnarray*}}
\newcommand{\ben}{\begin{enumerate}}
\newcommand{\een}{\end{enumerate}}
\newcommand{\e}{ { \mathbb{E}}}
\newcommand{\var}{ {\mathbb{V}\rm ar }}

\newcommand{\bX}{\bm{X}}

\newcommand{\bx}{\bm{x}}
\newcommand{\bU}{\bm{U}}
\newcommand{\bp}{\bm{p}}
\newcommand{\bq}{\bm{q}}

\newcommand{\convergeto}{ {\overset{d}{\longrightarrow \; }}}

\usepackage{authblk}

\begin{document}

\title{
Minimum Wasserstein distance estimator under covariate shift:
closed-form, super-efficiency and irregularity
}

\date{}

\author[1]{Junjun Lang}
\author[2]{Qiong Zhang\thanks{Corresponding author:  qiong.zhang@ruc.edu.cn}}
\author[1]{Yukun Liu\thanks{Corresponding author:  ykliu@sfs.ecnu.edu.cn}}

{\small
\affil[1]{ \small
		KLATASDS-MOE,  School of Statistics,
		East China Normal University,
		Shanghai, China}
\affil[2]{ \small
		Institute of Statistics and Big Data,
Renmin University of China, Beijing, China}
}

\maketitle

\bigskip	
\begin{abstract}
Covariate shift arises when covariate distributions differ between source and target populations while the conditional distribution of the response remains invariant, and it underlies problems in missing data and causal inference.
We propose a minimum Wasserstein distance estimation framework for inference under covariate shift that avoids explicit modeling of outcome regressions or importance weights.
The resulting W-estimator admits a closed-form expression and is numerically equivalent to the classical 1-nearest neighbor estimator, yielding a new optimal transport interpretation of nearest neighbor methods.
We establish root-$n$ asymptotic normality and show that the estimator is not asymptotically linear, leading to super-efficiency relative to the semiparametric efficient estimator under covariate shift in certain regimes, and uniformly in missing data problems.
Numerical simulations, along with an analysis of a rainfall dataset, underscore the exceptional performance of our W-estimator.
\end{abstract}

\noindent%

\section{Introduction}
Covariate shift is a fundamental challenge in machine learning and statistics, arising when the marginal distribution of the input variables (covariates or features) differs between the source (training) and target (test) populations, while the conditional distribution of the response given the covariates remains invariant~\citep{Shimodaira2000Improving, Quinonero-Candela2009}. 
Covariate shift occurs naturally in many applied problems. 
For example, in drug discovery, sample selection may depend solely on molecular characteristics such as amino acid sequences or chemical properties, leading to systematic differences between screened compounds and the broader chemical space. 
Similarly, in college admissions or hiring processes, historical selection mechanisms favoring certain demographic or academic attributes can induce substantial covariate shifts between previously selected cohorts and current applicant pools \citep{Jin2023Selection}.
If such distributional shifts are ignored, statistical inference and predictive models trained on the source data may yield biased estimates and degraded performance when applied to the target population.

The covariate shift framework underlies a large class of domain adaptation and transfer learning methods, whose central objective is to account for discrepancies in covariate distributions while preserving the conditional relationship between features and responses~\citep{Shimodaira2000Improving, He2024Covariate}.
\emph{A predominant approach for inference under covariate shift is based on importance weighting, in which source observations are reweighted by the ratio of the target and source covariate densities}. 
This idea was first introduced by~\citet{Shimodaira2000Improving}, and has since motivated a variety of methods for estimating importance weights, including approaches based on Kullback--Leibler divergence~\citep{sugiyama2007direct}, kernel mean matching ~\citep[KMM]{huang2006correcting, Gretton2008}, and least squares criteria~\citep{kanamori2009least}. 
Despite their popularity, importance weighting methods suffer from several well-known limitations. 
In addition to requiring the solution of potentially large-scale optimization problems, which can be computationally demanding, their performance may deteriorate substantially when the estimated weights take large values at a small number of training samples~\citep{pmlr-v202-segovia-martin23a}, resulting in unstable or high-variance estimators. \emph{An alternative line of work avoids explicit density ratio estimation by first estimating the conditional regression function using the source data and then integrating the resulting estimator with respect to the target covariate distribution.}
Estimators constructed in this manner are commonly referred to as plug-in estimators, as they replace the unknown conditional expectation in the target functional by its regression estimate learned from the source population.
However,~\citet{Yu2012607} showed that plug-in regression estimators are generally inferior to importance weighting methods based on kernel mean matching, in the sense that they achieve slower convergence rates.
In specific settings, such as estimation of the target population mean,~\citet{li2020robust} proposed a doubly robust estimator that combines nonparametric regression residuals with KMM-based importance weights; however, the resulting estimator does not achieve the parametric convergence rate.
Building on this idea,~\citet{Liu2023Argumented},~\citet{kato2023double}, and~\citet{yan2024transfer} extended doubly robust methodology to inference for more general regression parameters under covariate shift.
These approaches, however, typically rely on semiparametric modeling assumptions for either the importance weights or the regression function, and may involve substantial computational costs.

Although the notion of covariate shift was formalized in the machine learning literature, it captures a structural feature that is already implicit in several classical statistical problems.
In particular, missing data under the missing-at-random (MAR) assumption and causal inference under unconfoundedness can both be recast as covariate shift problems through appropriate theoretical reformulations.
In missing data settings, the MAR assumption~\citep{rubin1976inference} implies that the conditional distribution of the response given covariates is invariant between complete and incomplete samples, while the covariate distribution may differ due to selection.
Consequently, the distributions of the complete cases and the target population satisfy the covariate shift assumption, and naive inference based solely on complete cases may be biased.
Similarly, in causal inference within the potential outcome framework \citep{Neyman1923, Rubin1974estimating}, unconfoundedness implies a MAR condition for each potential outcome, reducing the estimation of treatment effects to a covariate shift problem.

Under MAR or unconfoundedness, a substantial literature has developed estimation procedures that can be broadly categorized into three classes.
\emph{The first class consists of outcome modeling approaches}, including regression and imputation-based methods~\citep{Hahn1998, chen2008semiparametric, ning2012comparison, Han2013417, Shao2019}, which construct complete datasets by imputing missing values, typically via regression models.
These methods rely on correct specification of the outcome model and often involve tuning parameters that can affect finite-sample performance~\citep{cheng1994nonparametric, wang2023statistical}.
\emph{The second class comprises weighting-based methods}, most notably inverse probability weighting (IPW)~\citep{horvitz1952generalization}, which reweights observations using the inverse of the propensity score, commonly estimated via nonparametric sieve methods~\citep{Hirano2003, chen2008semiparametric}.
While conceptually simple, IPW estimators are sensitive to propensity score misspecification and may suffer from high variance when the resulting weights are unstable~\citep{robins2007comment}.
\emph{The third class includes doubly robust methods}, such as augmented IPW (AIPW) estimators~\citep{Robins1994, RobinsRotnitzkyandzhao1995, robins1995semiparametric}, which combine outcome modeling and weighting to achieve consistency if either nuisance model is correctly specified, and semiparametric efficiency when both are correct. Again
simultaneous violation of both model assumptions can lead to biased or misleading estimates.
This problem can be mitigated by employing the double machine learning
~\citep{Chernozhukov2018}, which provides a general framework for implementing doubly robust estimators by using flexible machine learning methods to estimate nuisance functions while preserving root-$n$ consistency, albeit at the cost of increased computational complexity and tuning.

In this paper, we propose a novel minimum Wasserstein distance estimation framework for inference under covariate shift.
The proposed method applies broadly to covariate shift, missing data, and causal inference problems, all of which share a common structural form characterized by invariant conditional distributions and shifting covariate marginals.
\emph{Unlike existing approaches that rely on explicit modeling of either the outcome regression, the importance weights, or both, our framework directly aligns distributions across populations by minimizing a Wasserstein distance, thereby avoiding parametric or semiparametric nuisance modeling altogether.}
The Wasserstein distance, also known as the earth mover’s distance~\citep{Villani2003, Villani2009}, measures discrepancies between distributions by explicitly accounting for the geometry of the sample space.
In contrast to classical divergences such as Kullback–Leibler divergence or empirical likelihood, which are ill-defined or uninformative when the underlying distributions have non-overlapping supports, the Wasserstein distance remains well-defined and informative even when the source and target distributions are supported on different sets of covariate values—precisely the setting in which covariate shift is most severe.
Our main contributions are summarized as follows.

\begin{itemize}[leftmargin=*]
\item \textbf{A closed-form Wasserstein-based estimator without nuisance modeling}.
We propose a minimum Wasserstein distance estimator, referred to as the W-estimator, that admits a \emph{closed-form expression}. This stands in sharp contrast to general minimum Wasserstein distance estimators, which usually require solving large-scale optimization problems and do not admit explicit solutions~\citep{Villani2003, Villani2009, Peyre2019}.
Unlike standard covariate shift methods in both machine learning and statistics---which typically require estimating importance weights, outcome regressions, or both---the W-estimator avoids explicit nuisance function estimation entirely.

\item \textbf{Optimal transport interpretation of nearest neighbor methods}.
We show that the proposed W-estimator is numerically \emph{equivalent to the classical 1-nearest neighbor (1-NN) estimator} \citep{Fix1951, Cover1967, chen2000nearest, chen2001jackknife}.
This result provides a novel optimal transport interpretation of nearest neighbor methods, revealing them as implicit distribution-matching procedures under covariate shift.
Unlike standard $k$-NN estimators, which are typically analyzed as outcome-modeling tools, the W-estimator directly yields estimates of density ratios or importance weights and can be naturally integrated with regression and supervised learning procedures (see Section~\ref{sec:enhanced}).

\item \textbf{Root-$n$ inference without parametric or semiparametric assumptions}.
We establish that both the W-estimator and an enhanced version achieve root-$n$ convergence  and
are asymptotically normal under the covariate shift assumption.
This result contrasts with regression-based, importance weighting or IPW, and doubly robust estimators, whose asymptotic guarantees typically depend on correct specification or sufficiently fast convergence of nuisance estimators.

\item \textbf{Super-efficiency and asymptotic irregularity}.
We show that the proposed estimators can be more efficient than the semiparametric efficient estimator under covariate shift in certain regimes, and uniformly so in missing data problems.
Moreover, we demonstrate that the W-estimator is \emph{not asymptotically linear}, providing a theoretical explanation for its super-efficiency and distinguishing it fundamentally from influence-function-based estimators commonly used in covariate shift, missing data and causal inference.
\end{itemize}

The remainder of the paper is organized as follows. 
Section~\ref{sec-TL} introduces the minimum Wasserstein distance estimator in the covariate shift setting. 
Section~\ref{sec-mdp} extends the method to missing data problems. 
Simulation studies are presented in Section~\ref{sec-simu}, followed by a real data application in Section~\ref{sec-real}. 
Section~\ref{sec-discussion} concludes with discussions and remarks. 
All technical proofs are deferred to the supplementary material.

\section{W-estimator under covariate shift}
\label{sec-TL}
We introduce the proposed W-estimator in the context of covariate shift.  
Let $F_0(\bx, y)$ and $F_1(\bx, y)$ be  the source and target distribution
functions, respectively, of $(\bX, Y)$ on $\mathcal{X}\times \mathcal{Y}$, where $Y \in \mathcal{Y} \subset \mathbb{R}$ denotes a response and $ \bX \in \mathcal{X} \subset \mathbb{R}^d$ a $d$-dimensional vector of covariates.  
Suppose that $(\bX_1, Y_1), \ldots, (\bX_{n }, Y_{n })$ are $n$ independent and identically distributed (i.i.d.) observations from $F_{0}(\bx, y)$, and $(\tilde \bX_{1}, \tilde Y_{1}), \ldots, (\tilde \bX_{m}, \tilde  Y_{m})$ are
i.i.d. from $F_{1}(\bx, y)$, where the responses $\tilde Y_{j}$ are unavailable. 
In the following, we will always assume that the source data and target data are independent of each other.  
Let $F_k(\bx)$ denote the marginal distribution function of $\bX$ for the source ($k=0$) or target ($k= 1$) data.  
We assume that the source and target distributions satisfy the covariate shift assumption, that is  
\ba
\label{DRM} 
\frac{dF_1(\bx, y)}{dF_0(\bx, y)} =\frac{dF_1(\bx )}{dF_0(\bx )} = r(\bx),  \quad (\bx, y) \in \mathcal{X}\times \mathcal{Y}
\ea
for an unknown function $r(\cdot)$. 
For $k\in\{0,1\}$, let $\e_k$, $\mathbb{P}_{k}$ and $\var_{k}$ denote the  expectation, probability and variance operator, respectively, with respect to $F_k(\bx, y)$.    
For a pre-specified function  $g(\bx, y)$,  we wish to estimate $\theta = \e_1 \{g(\bX, Y)\} = \int  g(\bx, y) dF_{1}(\bx, y)$.

\subsection{Minimum Wasserstein distance estimator}
\label{sec:minimum-w-dist-estimate} 
As the source data are all available, a direct estimator for $F_{0}(\bx, y)$ 
is its empirical distribution  $F_{0n}(\bx, y) = \sum_{i=1}^{n } (1/n) I(\bX_i\leq \bx, Y_i\leq y)$.  
However, we can not estimate $F_{1}(\bx, y)$  by its empirical distribution because the target data are incomplete.  
Let $\bp = (p_1, \ldots, p_n)$ and $\mathcal{P} = \{ \bp = (p_1, \ldots, p_n) \mid  p_i\geq 0,~\sum_{i=1}^n  p_i=1\}$.
Under the covariate shift assumption~\eqref{DRM}, $F_{1}(\bx, y)$ is dominated by  $F_{0}(\bx, y)$, or equivalently, the support of $F_{1}(\bx, y)$ is a subset of that of $F_{0}(\bx, y)$.  
This motivates us to model $F_{1}(\bx, y)$ by a discrete distribution function
with support points being the support of $F_{0n}(\bx, y)$ or equivalently 
the $n$  observations from $F_0(\bx, y)$, that is, 
\ba \label{discrete-F}
F_1(\bx, y; \bp)=  \sum_{i=1}^{n } p_i I(\bX_i\leq \bx, Y_i\leq y), \quad \bp \in \mathcal{P}.
\ea 
When one constructs a reasonable estimator $\hat \bp$ for $\bp$, we immediately have an estimator  $F_1(\bx, y; \hat \bp)$ for $F_{1}(\bx, y)$, and accordingly a reasonable estimator $\int  g(\bx, y) dF_1(\bx, y; \hat \bp)$ for $\theta$. 
Therefore, our problem reduces to estimating $\bp$.

Note that under model \eqref{discrete-F},  the marginal distribution $F_{1}(\bx)$ of $\bX$ is
\bas
F_1(\bx; \bp )=F_1(\bx, \infty; \bp)=  \sum_{i=1}^{n } p_i I(\bX_i\leq \bx), \quad \bp\in \mathcal{P}.
\eas
In the meantime, the empirical distribution $F_{1m}(\bx ) =  \sum_{i=1}^{m } (1/m) I(\tilde\bX_{i} \leq \bx)$ of the observed target data is also a reasonable estimator of   $F_{1}(\bx)$. 
A natural estimator of $\bp$ can be obtained by minimizing a distance/divergence between $F_1 (\cdot; \bp )$ and $F_{1m}(\cdot)$ over $F_1(\cdot; \bp)\in  \mathscr{F}_{\bX} = \{F_1 (\bx; \bp ):\bp \in \mathcal{P}\}$.   
There are many distances or divergences that are used to quantify the closeness of two distributions.
Popular examples include   Kullback-Leibler divergence \citep{Kullback1951},  Jensen-Shannon divergence \citep{Lin1991}, Hellinger distrance \citep{Rudolf1977},
Pearson's chisquare divergence \citep{Pearson1900},  Neyman's chisquare divergence \citep{Broniatowski2006},
 empirical likelihood  \citep{Owen1988,Qin1994}, Euclidean likelihood \citep{Owen1991},  Exponential tilting likelihood \citep{Efron1981},  and  more general  Cressie-Read divergences \citep{Cressie1984}. Unfortunately, all these distances or divergences fail to work here because  they all require  the two distributions under study to have common support points but  $F_1(\cdot; \bp)$ and $F_{1m}(\cdot)$ do not.

This problem can be circumvented  by  the  Wasserstein distance, also known as the earth mover’s distance \citep{Villani2003,Villani2009}. The Wasserstein distance incorporates the geometry of the underlying space by considering the cost of transporting probability mass. As a result, it is particularly powerful when comparing distributions with non-overlapping or distant supports, where the aforementioned divergence measures may be infinite or uninformative.  
Let   $\Pi(F, G)$ denote the set of joint distribution functions with marginal distributions being $F $ and $G$.
For any $q\geq 1$,  the  $q$-Wasserstein distance between $F$ and $G$ is
\bas
W_q(F, G)
=
\left\{  \min_{\Gamma  \in \Pi(F, G)}   \e_{(\bU_{1}, \bU_{2})\sim \Gamma} \|\bU_{1} - \bU_{2}\|_q^q \right\}^{1/q}, 
\eas
where  $\|\cdot\|_q$ denote the  $L_q$ norm   for a vector in $\mathbb{R}^{d}$.
When $q=1$,  the 1-Wasserstein  metric is  known as the earth mover distance.
As  $F_1(\cdot; \bp)$ and  $ F_{1m}$ are discrete distribution functions,  we  have a better expression of
their   Wasserstein distance, i.e. 
\bas
W_q(F_1(\cdot; \bp), F_{1m}(\cdot))
= 
\left[ \min
\left\{  \sum_{i=1}^n \sum_{j=1}^m \pi_{ij}  \|\bX_i - \tilde\bX_{j}  \|_q^q\ \bigg| \  \pi_{ij} \geq 0, \;
\sum_{i=1}^n \pi_{ij} = \frac{1}{m}, \;  \sum_{j=1}^m \pi_{ij} = p_i   \right\} \right]^{1/q}.
\eas
Hereafter we focus on the case  $q=2$ although our method works for any fixed  $q$
and write $\|\cdot\|_2$ as  $\|\cdot\|$  for short.

We   propose to estimate $F_1(\cdot; \bp)$  or equivalently $\bp$ by minimizing $W_2(F_1(\cdot; \bp), F_{1m}(\cdot))$, 
i.e.
\ba \label{mini-hat-p}
  \hat\bp := (\hat p_1, \ldots, \hat p_n) =    \arg\min_{ \bp  \in  \mathcal{P}}    W_2(F_1(\cdot; \bp), F_{1m}(\cdot)).
\ea
Although the Wasserstein distance  $ W_2(F_1(\cdot; \bp), F_{1m}(\cdot))$  has no closed-form in general,   
we surprisingly find that  the minimizer $\hat\bp $ does.
For  $1\leq i\leq n, 1\leq j\leq m$, let
  $e_{ij} = \|\bX_{i } - \tilde \bX_{j}\|$.
For each $1\leq j\leq m$, let 
$\mathcal{S}_{j} = \{ k \mid   e_{kj} = \min_{1\leq l\leq n}  e_{lj} \} $, which is 
 the set of indice   $  k $ for which $\bX_k$  are the closest to $\tilde \bX_j$
 among $ \{ \bX_1, \ldots, \bX_n\}$.  Let $I(A)$ be the indicator function of the event $A$.  

\begin{theorem}
	\label{thm-close-form} 
(i)The vector  $\hat\bp = (\hat p_1, \ldots, \hat p_n)$  
in \eqref{mini-hat-p}  must  satisfy 
$
\hat p_i = \sum_{j=1}^m   \hat \pi_{ij}
$
for 
$1\leq i\leq n$, 
where    
\( 
\hat \pi_{ij}
=
 \alpha_{ij} I(   i  \in  \mathcal{S}_{j} )/m 
 \)
and $\alpha_{ij}$ satisfy 
 $\alpha_{ij}\geq 0$ and $\sum_{l\in \mathcal{S}_{j}} \alpha_{lj} = 1$. 
 (ii) 
 If $F_0(\bx)$  is  absolutely  continuous,   
 then  with probability one,  
$
\hat \pi_{ij}
=   (1/m) \prod_{k=1}^n I(e_{ij}\leq e_{kj})
$, $  1\leq j\leq m$,  
and  the corresponding $\hat\bp  $  is the unique minimizer of $  W_2(F_1(\cdot; \bp), F_{1m}(\cdot))$ 
 over $\mathcal{P}$. 
 
\end{theorem}

Hereafter we always assume that $F_0(\bx)$  is  absolutely  continuous so that 
with probability one, the set $\mathcal{S}_{j}$ contains only one element  
 and     $ e_{1j}, \ldots,  e_{nj}$ are all different from each other for all $1\leq j\leq m$. 
Therefore  
\ba
\label{definition-pi}
\hat p_i = \sum_{j=1}^m   \hat \pi_{ij}= \frac{1}{m}  \sum_{j=1}^m  \prod_{k=1}^n I(e_{ij}\leq e_{kj}), \quad
1\leq i\leq n.
\ea
Once $\hat \bp = (\hat{p}_{1}, \cdots, \hat{p}_{n})$ is obtained,  a natural estimate of $F_1(\bx, y)$ is 
$
 F_1(\bx, y; \hat \bp) = \sum_{i=1}^n  \hat p_i I(\bX_i\leq \bx, Y_i \leq y).
$
Accordingly,  the proposed W-estimator for     $\theta  = \int g(\bx, y) dF_{1}(\bx, y)$  is
\ba
\label{eq:theta-hat-w-vanilla}
\hat \theta
 =  \int g(\bx, y) d  F_1(\bx, y; \hat \bp) 
=  \sum_{i=1}^n g(\bX_i, Y_i) \hat p_i
=  \frac{1}{m}   \sum_{j=1}^m   \sum_{i=1}^n g(\bX_i, Y_i)   \prod_{k=1}^n I(e_{ij}\leq e_{kj}).
\ea
It can be verified that   $\hat{\theta}$ is equal to the popular 1-nearest neighbor imputation estimator (1-NNI) in survey sampling \citep{chen2000nearest, chen2001jackknife, yang2020asymptotic}.   
Unlike these works, our method can produce  direct estimates for importance weights  or density ratios at the source data points $r(\bX_i)$,  i.e. 
\bas
\widehat{r(\bX_i)} =  \frac{dF_1(\bX_i, Y_i; \hat \bp)}{dF_{0n}(\bX_i, Y_i)} =  n \hat p_i, \quad 1\leq i \leq n. 
\eas

\subsection{Asymptotic properties of the W-estimator}

Let  $N=n+m$. For $s=1,2,4$, let $g_{s}(\bx) = \e\left\{g^{s}(\bx, Y)\mid \bX = \bx\right\}$. For any $\bx\in\mathcal{X}$ and $t\geq 0$, let $G_{\bx}(t) = \mathbb{P}(\|\bX_1 - \bx \|\geq t)$. Furthermore, for any $\bx_{1}, \bx_{2}\in\mathcal{X}$ and $t_{1}, t_{2}\geq 0$, let $G_{\bx_{1}, \bx_{2}}(t_{1}, t_{2}) = \mathbb{P}\left(\|\bX_{1} - \bx_{1}\|\geq t_{1}, \|\bX_{1} - \bx_{2}\|\geq t_{2}\right)$. For any function $h(\cdot)$, possibly vector-valued, let $\|h\|_{\infty} = \sup_{\bx\in\mathcal{X}}\|h(\bx)\|$, $\nabla_{\bx}h(\bx) = \partial h(\bx)/\partial \bx$ and $\nabla^{2}_{\bx}h(\bx) = \partial^{2} h(\bx)/\partial \bx\partial \bx^{\top}$. 
We use $\lambda_{\max}(A)$ to denote the largest eigenvalue of matrix $A$ and use $\convergeto$ to denote ``convergence in distribution''. 

 \begin{assumption}
	\label{ass-uniqueness}
	Suppose that   $F_0(\bx)$  is  absolutely  continuous
	and $F_1(\bx, y)$ is  absolutely continuous with respect to $F_0(\bx, y)$.
\end{assumption} 

\begin{assumption} \label{assump-bounded-derivative} There exist positive constants $M_{1}, M_{2}, M_{3}$ such that: (i) $\|g(\cdot)\|_{\infty}\leq M_{1}$.
	(ii) 
	$ \| \nabla_{\bx} g_1(\cdot)\|_{\infty} +  \| \nabla_{\bx} g_2(\cdot)\|_{\infty}+\| \nabla_{\bx} g_4(\cdot)\|_{\infty} \leq M_{2}$. (iii) $\|\lambda_{\max}(\nabla_{\bx}^{2}g_{1}(\cdot))\|_{\infty}+\|\lambda_{\max}(\nabla_{\bx}^{2}g_{2}(\cdot))\|_{\infty}\leq M_{3}$.
\end{assumption}

\begin{assumption}
	\label{assump-covariate-distribution} (i)
$n \e[  \|\bX_1 - \tilde \bX_1 \|^2  \{ G_{\tilde \bX_1} (\|\bX_1 - \tilde \bX_1 \|    ) \}^{n-1} ]  = o(n^{-1})$. (ii) $n \e[ \{ G_{\tilde \bX_1, \tilde \bX_2} (\|\bX_1 - \tilde \bX_1 \|  ,
\|\bX_1 - \tilde \bX_2 \|    ) \}^{n-1} ]  = O(n^{-1})$. (iii) $n \e[    \|\bX_{1} - \tilde{\bX}_{1}\|\{ G_{\tilde \bX_1, \tilde \bX_2} (\|\bX_1 - \tilde \bX_1 \|  ,
\|\bX_1 - \tilde \bX_2 \|    ) \}^{n-1} ]  = o(n^{-1})$.
\end{assumption}

Assumption \ref{assump-bounded-derivative}  is imposed for technique convenience and may be relaxed.  
 Assumption \ref{assump-covariate-distribution}  imposes several restrictions on the 
source and target covariate distributions.  Although nontrivial,  
the conditions in Assumption \ref{assump-covariate-distribution}   are not too restrictive because 
they  are satisfied when  $\bX_{1}$ and $\tilde \bX_{1}$ are uniformly distributed on the interval $[0,1]$. See Section 5 in the supplementary material.

\begin{theorem}
\label{thm-asy-theta-hat-TL}  
Under Assumptions \ref{ass-uniqueness}--\ref{assump-covariate-distribution} and model \eqref{DRM}, if   $m \rightarrow\infty$ as $n\rightarrow\infty$ and $m   = O(n)$, then
$
\sqrt{m} (\hat{\theta} - \theta )\stackrel{d}{\longrightarrow}\mathcal{N}(0,\sigma^2)
$
as $n\rightarrow\infty$, where
$ \sigma^{2} = \var_1 \{g(\bX, Y) \}$.
\end{theorem}

A critical challenge in our proof of Theorem \ref{thm-asy-theta-hat-TL}  is that  the classical linear approximation technique fails to work here.  We crack the nut by  transforming our estimator  to a martingale and then applying 
the central limit theorem for  martingales \citep[Theorem 3.2]{HALL198051}. Theorem \ref{thm-asy-theta-hat-TL}   indicates that our W-estimator  has a  root-$n$ convergence rate and asymptotic normality,
whereas  the NNI estimator usually has a rate of lower than root-$n$ and  its asymptotic normality usually involves a bias correction term \citep{Abadie2006, Abadie2012, Abadie2016}.   Under covariate shift, \cite{Portier2024} introduced an estimator for target population characteristics based on Nearest Neighbor Sampling.  When the 1-nearest neighborhood is used,    this estimator is equal to  our W-estimator $\hat{\theta}$.  \cite{Portier2024}   demonstrated that the convergence rates of their estimators are of the order of 
$m^{-1/2}+n^{-1/d}$, however  they did not establish its asymptotic normality.

Given its asymptotic normality, is the proposed W-estimator $\hat \theta$ asymptotically semiparametric efficient? 
The semiparametric efficiency theory on  the population mean estimation has  been extensively studied on missing data problems \citep{tsiatis2006semiparametric},
however there seems no such a theory for the estimation of $\theta$ in the context of covariate shift.  Lemma \ref{lemma-efficiency-lower-bound}  presents its semiparametric efficiency lower bound, i.e., the asymptotic variance lower bound for all {\it regular and asymptotically linear} (RAL) estimators of $\theta$.  See \cite{newey1990semiparametric} and Chapter 3 of \cite{tsiatis2006semiparametric} for the definition of an RAL estimator.  
\begin{lemma}
\label{lemma-efficiency-lower-bound}
Suppose that  model \eqref{DRM} is satisfied and  $\lim_{n\rightarrow\infty}n/N =\eta\in(0,1)$. 
Then  the semiparametric efficiency lower bound for  the estimation of $\theta$ is
	\bas
	\sigma_{\rm eff}^2 = \frac{1}{1-\eta}\var_{1}(g_{1}(\bX))+\frac{1}{\eta}\e_{0}\left[r^{2}(\bX)\var(g(\bX, Y)\mid \bX)\right].
	\eas
\end{lemma}

The  lower bound $\sigma_{\rm eff}^2 $ is  obtained  with respect to the whole sample size $N$. Clearly, by Theorem \ref{thm-asy-theta-hat-TL},   $\sqrt{N}(\hat{\theta} - \theta)\stackrel{d}{\rightarrow}\mathcal{N}(0, \sigma^{2}_{\text{W}})$, where $\sigma^{2}_{\text{W}} = (1-\eta)^{-1}\sigma^{2}$.
Because
$
\sigma^{2} = \var_1(g_{1}(\bX)) + \e_{1}\left\{\var\left(g(\bX, Y)\mid \bX\right)\right\}
$
and $\e_{0}\left[ r^{2}(\bX)\var\left(g(\bX, Y)\mid \bX\right) \right] 
= \e_{1}\left\{r(\bX)\var(g(\bX, Y)\mid \bX)\right\},$ we have
\bas
\sigma_{\rm eff}^2  - \sigma^{2}_{\text{W}} = \e_{1}\left[\var\left(g(\bX, Y)\mid \bX\right)\left\{\frac{1}{\eta}r(\bX) - \frac{1}{1-\eta}\right\}\right].
\eas
This implies that the proposed W-estimator
is more efficient than a semiparametric efficient estimator
when the right hand side of the above is greater than $0$. The following Corollary gives a special case for our estimator to be super efficient.

\begin{corollary}\label{super-efficiency-TL}
Under the assumptions in Theorem \ref{thm-asy-theta-hat-TL} and Lemma \ref{lemma-efficiency-lower-bound}, further assume that $\mathbb{P}_{0}(r(\bX)\neq 1)>0$, $\eta \leq 1/2$ and  $\var (g(\bX, Y)\mid \bX ) = \sigma_{1}^2$ for some $\sigma_{1}^{2}> 0$, then  the proposed W-estimator $\hat{\theta}$ is strictly more efficient than a semiparametric efficient estimator for $\theta$, or  asymptotically  super efficient.
\end{corollary}

Our estimator constitutes a novel super-efficient estimator in the statistical literature, alongside the well-established Hodges estimator \citep{Bickel1998} and the Hodges-Le Cam estimator \citep{Kale1985}.
The result in Corollary \ref{super-efficiency-TL} seems weird as it   violates
the semiparametric efficiency lower bound.
This uncommon phenomenon can be explained by
the fact that  the proposed W-estimator is not
an RAL  estimator.

\begin{prop} \label{prop-not-ral-TL}
Suppose that  the assumptions in Theorem \ref{thm-asy-theta-hat-TL} and Lemma \ref{lemma-efficiency-lower-bound} are satisfied, and $\sigma^{2}>0$. Then  it never holds that
\bas
\sqrt{N}(\hat{\theta}-\theta)  = \frac{1}{\sqrt{N}} \left\{
\sum\limits_{i=1}^{n} L_1(\bX_i, Y_i) + \sum\limits_{j=1}^{m} L_2(\tilde \bX_{j}) \right\}
+o_{p}(1) 
\eas
  as $n\rightarrow\infty$,  
for any functions $L_{1}(\bx, y)$ and $L_{2}(\bx)$ satisfying (i) $\e\{L_{1}(\bX_{1}, Y_{1})\} = \e\{L_{2}(\tilde \bX_{1})\} = 0$, (ii) $\e\{L_{2}^{2}(\tilde \bX_{1})\}<\infty$, (iii) $\bx\mapsto \e\{L_{1}(\bx, Y)\mid \bX=\bx\} $ is a differentiable and uniformly bounded function and (iv) $\bx\mapsto \e\left\{g(\bx, Y)L_{1}(\bx, Y)\mid \bX = \bx\right\}$ is a differentiable function. 
\end{prop}

Proposition \ref{prop-not-ral-TL} reveals that, in a general context, the proposed W-estimator $\hat{\theta}$ is not an asymptotically linear estimator, and consequently  not an RAL estimator. The semiparametric efficiency lower bound $ \sigma_{\rm eff}^2 $
is the minimum  of the asymptotic variances of all
RAL estimators of  $\theta$.
Since our W-estimator
$\hat{\theta}$ is not an RAL   estimator  of $\theta$,
 it is possible for $\hat{\theta}$
to have an asymptotic variance smaller than the semiparametric efficiency lower bound $ \sigma_{\rm eff}^2 $.

When making inference about $\theta$ based on Theorem \ref{thm-asy-theta-hat-TL},
it is necessary to construct a reasonable estimator for  its asymptotic variance $\sigma^2$. 
We propose to estimate   $\sigma^2$  by
\ba
\label{eq:hat-var}
\hat \sigma^2 =    \sum_{i=1}^n  \hat p_i   \{ g(\bX_i, Y_i)\}^2 -   \hat{\theta}^{2} =   
   \frac{1}{m}   \sum_{j=1}^m \sum_{i=1}^n \{ g(\bX_i, Y_i)\}^2   \prod_{k=1}^n I(e_{ij}\leq e_{kj})
   - \hat{\theta}^{2}.
\ea
This estimator is clearly consistent by applying Theorem \ref{thm-asy-theta-hat-TL}.

\begin{prop} \label{prop-var-est}
Under Assumptions \ref{ass-uniqueness}--\ref{assump-covariate-distribution}  and model \eqref{DRM}, 
it holds that  $\hat \sigma^2 = \sigma^2 + o_p(1)$ as $n$ and $m$ go to infinity.
\end{prop}

\subsection{Enhanced W-estimation}  
\label{sec:enhanced}

The proposed W-estimator is  so flexible that it can incorporate the state-of-art statistical regression 
and machine learning algorithms to enhance performance.  
Given any nonrandom function $\zeta(\bx)$,  let  $  \epsilon = g(\bX, Y) - \zeta(\bX)$ and 
  $F_{\epsilon k}(\bx, \epsilon)$ denote the joint distribution function of  $(\bX, \epsilon)$
when the corresponding  $(\bX, Y)$ follows $F_k(\bx, y)$,  $k=0, 1$.
It can be directly verified that if  $F_0(\bx, y)$ and $F_1(\bx, y)$ satisfy  model \eqref{DRM}, then
so do $F_{\epsilon 0}(\bx, \epsilon)$ and  $F_{\epsilon 1}(\bx, \epsilon)$.
 Based on the data  $\{(\bX_i, \epsilon_i),  1\leq i\leq n\}$ and
 $\{\tilde\bX_{1}, \ldots, \tilde\bX_{m}\}$,
we can apply the standard W-estimation method to construct a W-estimator
$\hat F_{\epsilon 1}(\bx, \epsilon) = \sum_{i=1}^n \hat p_i I(\epsilon_i\leq \epsilon, \bX_i\leq \bx)$ for   $F_{\epsilon 1}(\bx, \epsilon)$, where $\hat p_i$ are exactly those defined in \eqref{definition-pi}. 
Then,  with the fixed $\zeta(\bx)$, an enhanced W-estimator of
\bas
\theta &=& \int g(\bx, y) dF_1(\bx, y)= \int \zeta(\bx) dF_1(\bx, y)+  \int  \epsilon  dF_{\epsilon 1}(\bx, \epsilon)
\eas
 is
\bas
\tilde{\theta}_{\rm en}   
=   \frac{1}{m}\sum_{j=1}^m \zeta(\tilde\bX_{j}) +   \sum_{i=1}^n \hat p_i  \{g(\bX_i, Y_i) - \zeta(\bX_i)\}.
\eas
This enhanced estimation  procedure works for any pre-specified nonrandom function  $\zeta(\bx)$,
e.g.   $\e\{ g(\bX, Y)|\bX=\bx \}$.
In practice, we may replace  $\zeta(\bx)$ by a  statistical regression  or
  machine learning model such as linear regression,
 neural network, random  forests etc.  Please note that in this situation, we take  $g(\bX, Y)$ instead of $Y$ as a response.
To avoid over fitting, we shall use the  data-splitting technique
to construct an estimator of $\theta$ together with the proposed W-estimation method. 
Algorithm \ref{algorithm-data-splitting} presents a  detailed implementation of  the data-splitting based Enhanced W-estimator. 

\begin{algorithm}[H]
\label{algorithm-data-splitting}
  \SetAlgoLined
  \KwData{Source data $\{(\bX_{i}, Y_{i}):  i \in  \mathcal{I}^s \}$ with $ \mathcal{I}^s = \{1,2, \ldots, n\}$; 
  Target data $\{ \tilde \bX_{j}:  j \in  \mathcal{I}^t\}$ with  $ \mathcal{I}^t= \{ 1,2, \ldots, m\}$;  a regression algorithm $\mathcal{A}$}
  \KwResult{$\hat{\theta}_{\rm en} $}
  
Randomly split  $ \mathcal{I}^s$  into two halves,   $\mathcal{I}_1^s$ and $\mathcal{I}_2^s$  with  sample sizes $n_1$ 
and $n_2$,   respectively.   
Randomly split   $  \mathcal{I}^t$  into two halves,   $\mathcal{I}_1^t$, and $\mathcal{I}_2^t$  with  sample sizes   $m_1$, and $m_2$, respectively.

Fit   regression functions $\hat{\zeta}^{(i)}$ with  algorithm $\mathcal{A}$  based on 
  $\{(\bX_{j}, Y_{j}):  j \in  \mathcal{I}_{i}^{s} \}$ for $i=1, 2$.

For $i=1,2$, calculate  
$ 
\hat{p}_{l}^{(i)} =  (1/m_{i})  \sum_{j\in\mathcal{I}_{i}^{t}}  \prod_{k\in\mathcal{I}_{i}^{s}}I(e_{lj}\leq e_{kj}), \quad
l\in\mathcal{I}_{i}^{s}.
$

Calculate the enhanced W-estimates for each pair of datasets, 
\bas
\hat{\theta}^{(1)}_{\rm en} &=& \frac{1}{m_{1}}\sum\limits_{j\in\mathcal{I}_{1}^{t}}\hat{\zeta}^{(2)}(\tilde\bX_{j})+\sum\limits_{i\in\mathcal{I}_{1}^{s}}\hat{p}_{i}^{(1)}\{g(\bX_{i}, Y_{i}) - \hat{\zeta}^{(2)}(\bX_{i})\},\\
\hat{\theta}^{(2)}_{\rm en} &=& \frac{1}{m_{2}}\sum\limits_{j\in\mathcal{I}_{2}^{t}}\hat{\zeta}^{(1)}(\tilde\bX_{j})+\sum\limits_{i\in\mathcal{I}_{2}^{s}}\hat{p}_{i}^{(2)}\{g(\bX_{i}, Y_{i}) - \hat{\zeta}^{(1)}(\bX_{i})\}.
\eas

Output
$
\hat{\theta}_{\rm en} = \frac{n_{1}+m_{1}}{n+m}\hat{\theta}^{(1)}_{\rm en}+\frac{n_{2}+m_{2}}{n+m}\hat{\theta}^{(2)}_{\rm en}.
$ 
\caption{Calculation of data-spiltting based W-estimator} 
\end{algorithm}

Theoretically we find that  the Enhanced W-estimator possesses the same asymptotic distribution 
as the original W-estimator,   regardless of the regression algorithm $\mathcal{A}$ chosen.  
For this result to hold,  we  impose additional conditions.   
 
\begin{assumption}
    \label{ass-enhanced}
    (i) $m_{1} = m_{1,n}\rightarrow\infty$ and $m_{2} = m_{2,n}\rightarrow\infty$. (ii) $m = O(n)$ and $\lim_{n\rightarrow\infty} m_{1,n}/m = \lim_{n\rightarrow\infty}n_{1}/n = \tau$ for some $\tau\in(0,1)$. (iii) $\e\{\|\bX_{1} - \tilde \bX_{1}\|\mid \tilde \bX_{1}\}<\infty$ almost surely. (iv) $\e\{\|\nabla_{\bx}\hat \zeta^{(2)}(\cdot)\|_{\infty}\}+\e\{\|\nabla_{\bx}\hat \zeta^{(1)}(\cdot)\|_{\infty}\}<\infty$.
\end{assumption}

\begin{theorem}
	\label{thm-asymptotic-enhanced}
Suppose that the Assumptions \ref{ass-uniqueness}--\ref{ass-enhanced} and model \eqref{DRM} are satisfied. Then $
	\sqrt{m} (\hat{\theta}_{\rm en} - \theta )\stackrel{d}{\longrightarrow}\mathcal{N}(0,\sigma^2)
	$
	as $n\rightarrow\infty$.
\end{theorem}
Theorem \ref{thm-asymptotic-enhanced} indicates that under certain conditions, 
the enhanced W-estimator
has the same asymptotically normal distribution,
and therefore also possesses  super-efficiency.

\section{Extension to  missing data problems}
\label{sec-mdp}

The proposed W-estimator not only works for covariate shift but 
 is  also applicable to  missing data problems and causal inference,
although here we  consider missing data problems only.  
Suppose that $Y$  is subject to missingness and let  $D$  denote the
 missingness indicator, where  $D=1$ if and only if  $Y$ is observed, and 0 otherwise.
Suppose that    $(\bX_i, Y_i, D_i),  1\leq i\leq N $, 
 are iid copies of  $(\bX, Y, D)$.   We wish to estimate 
   $\mu = \e\{g(\bX, Y)\}$   for a pre-specified function $g$ 
   under the MAR assumption, i.e.  $\mathbb{P}(D=1|Y, \bX) = \mathbb{P}(D=1|\bX)$. 
 
 For $k\in\{0,1\}$, let $F_k(\bx, y)$ and $F_{k}(\bx)$ denote the joint distribution function of $(\bX, Y)$ and marginal distribution function of $\bX$ given  $D=1-k$, respectively. 
Denote   $\eta = \e(D)$,  $n = \sum_{i=1}^N  I(D_i=1)$, $m=N-n$, $[N] = \{1, \ldots,  N\}$, 
$\mathcal{N}_{1} = \{i: D_{i} = 1, i\in[N]\}$ and $\mathcal{N}_{0} = \{i: D_{i} = 0, i\in[N]\}$.

\subsection{Minimum Wasserstein distance estimator}

The MAR assumption implies the  covariate shift assumption in  \eqref{DRM}  holds.  
Similar to the covariate shift case, we model   $F_{1}(\bx, y)$ by
 \bas
 \tilde F_1(\bx, y; {\bq}) = \sum\limits_{i\in\mathcal{N}_{1}}q_{i}I(\bX_{i}\leq \bx, Y_{i}\leq y), \quad \bq\in\mathcal{Q},
 \eas
 where  $\bq = (q_{j}, j \in \mathcal{N}_{1})$ and  $\mathcal{Q} =\{\bq \mid q_{i}\geq 0, ~i\in\mathcal{N}_{1},~\sum_{i\in\mathcal{N}_{1}}q_{i} = 1\}$. 
This immediately leads to  a model for marginal distribution $F_{1}(\bx)$,
 \bas
\tilde F_1(\bx; {\bq} ) =\tilde F_1(\bx, \infty; {\bq})=  \sum_{i\in\mathcal{N}_{1}} q_i I(\bX_i\leq \bx),
  \quad \bq \in \mathcal{Q}.
\eas
Since   the empirical distribution $\tilde F_{1m}(\bx) = m^{-1}\sum_{i\in\mathcal{N}_{0}}I(\bX_{i}\leq \bx)$
is a natural estimator of the same  $F_{1}(\bx)$,   we  propose to estimate  $F_{1}(\bx)$ by  
$\tilde F_1(\bx; \hat{\bq})$, where 
 \bas
\hat \bq  \equiv (\hat q_{j}, j \in \mathcal{N}_{1})
=    \arg\min_{\bq \in \mathcal{Q}}    W_2(\tilde F_1(\cdot; {\bq} ), \tilde F_{1m}(\cdot)). 
\eas
  Similar to Theorem \ref{thm-close-form},  it can be found that   
 \bas
 \hat{q}_{i} = \frac{1}{m}\sum\limits_{j\in\mathcal{N}_{0}}\prod_{k\in\mathcal{N}_{1}}I(\tilde e_{ij}\leq \tilde e_{kj}), 
 \quad  i\in\mathcal{N}_{1},  
 \eas
 where $\tilde{e}_{tq} = \|\bX_{t} - \bX_{q}\|$ for $t, q\in[N]$.

 Therefore, the proposed W-estimator for $F_{1}(\bx, y)$ is 
 $
 \tilde F_1(\bx, y; {\hat \bq}) = \sum_{i\in\mathcal{N}_{1}} \hat q_{i}I(\bX_{i}\leq \bx, Y_{i}\leq y), 
$
 and accordingly 
 the proposed W-estimator for  
  $\mu_{1} =\e_{1}\{ g(\bX, Y)\} =\e\{ g(\bX, Y)|D=0\}$ is
 \bas
 \hat{\mu}_{1} 
 =
 \sum_{i\in\mathcal{N}_{1}}\hat{q}_{i}g(\bX_{i}, Y_{i})
 =
 \frac{1}{m} \sum\limits_{j\in\mathcal{N}_{0}}\sum\limits_{i\in\mathcal{N}_{1}}g(\bX_{i}, Y_{i})\prod_{k\in\mathcal{N}_{1}}I(\tilde{e}_{ij}\leq \tilde{e}_{kj}).
 \eas
In the meantime, the moment estimator of $\mu_{0} =\e_{0}\{ g(\bX, Y)\}  = \e\{ g(\bX, Y)|D=1\}$ 
is  $\hat{\mu}_{0} =  (1/n)\sum_{i\in\mathcal{N}_{1}}g(\bX_{i}, Y_{i})$. 
Because   $\mu = \eta\mu_{0}+ (1-\eta)\mu_{1}$ 
and the moment estimator of $\eta$ is  $\hat \eta = n/N$, 
 we propose to estimate  $\mu$  by 
 \bas
 \hat{\mu} = (1-\hat \eta)\hat{\mu}_{1}+ \hat\eta\hat{\mu}_{0} 
 = \frac{1}{N} \left\{ 
 \sum_{i\in\mathcal{N}_{1}}g(\bX_{i}, Y_{i})
 +
 \sum\limits_{j\in\mathcal{N}_{0}}\sum\limits_{i\in\mathcal{N}_{1}}g(\bX_{i}, Y_{i})\prod_{k\in\mathcal{N}_{1}}I(\tilde{e}_{ij}\leq \tilde{e}_{kj}) 
 \right\}, 
 \eas
 which is still called  a W-estimator.

\subsection{Super-efficiency and enhanced W-estimation}
Let $\tilde \bX$ and $\bar{\bX}$ denote independent realization 
from the distributions of $\bX\mid D = 0$ and $\bX\mid D=1$, respectively. 
Define $\tilde{G}_{\bx}(t) = \mathbb{P}\left(\|\bar{\bX} - \bx\|\geq t\right)$ 
and     $\tilde G_{\bx_{1}, \bx_{2}}(t_{1}, t_{2}) = \mathbb{P}(\|\bar{\bX} -\bx_{1}\|\geq t_{1}, \|\bar{\bX} - \bx_{2}\|\geq t_{2})$ for any $\bx, \bx_{1}, \bx_{2}\in\mathcal{X}$ and $t, t_{1}, t_{2}\geq 0$. 
When studying the asymptotic properties of the proposed W-estimator   $\hat \mu$, 
we postulate Assumptions \ref{assump-bounded-derivative-missing} and \ref{assump-covariate-distribution-missing},
which  are  modified versions of Assumptions \ref{assump-bounded-derivative} and  \ref{assump-covariate-distribution}, respectively, for missing data problems.

\begin{assumption} \label{assump-bounded-derivative-missing} 
(i) Assumption \ref{assump-bounded-derivative}(i) and (iii) are satisfied.  (ii) There exists a $\tilde M_{2}\geq0$, such that
	$ \| \nabla_{\bx} g_1(\cdot)\|_{\infty} +  \| \nabla_{\bx} g_2(\cdot)\|_{\infty} \leq \tilde M_{2}$. 
\end{assumption}
\begin{assumption}
\label{assump-covariate-distribution-missing} As $k$ is large,  
(i) $k \e[  \|\bar{\bX} - \tilde \bX \|^2  \{ \tilde G_{\tilde \bX} (\|\bar{\bX} - \tilde \bX \|    ) \}^{k-1} ]  = o(k^{-1})$.  (ii) $k \e[ \{ \tilde G_{\tilde \bX_1, \tilde \bX_2} (\|\bar{\bX} - \tilde \bX_1 \|  ,
\|\bar{\bX} - \tilde \bX_2 \|    ) \}^{k-1} ]  = O(k^{-1})$. 
(iii) $k \e[    \|\bar{\bX} - \tilde{\bX}_{1}\|\{ \tilde G_{\tilde \bX_1, \tilde \bX_2} (\|\bar{\bX} - \tilde \bX_1 \|  ,
\|\bar{\bX} - \tilde \bX_2 \|    ) \}^{k-1} ]  = o(k^{-1})$.
\end{assumption}

We show that the proposed W-estimator $\hat{\mu}$ retains an asymptotically normal distribution in the setting of missing data problems, as stated in Theorem \ref{thm-asy-mu-hat-MDP}. Similar to the covariate shift case, conventional linear approximation methods remain ineffective for establishing the asymptotic normality of $\hat{\mu}$. To overcome this difficulty, we employ the conditional central limit theorem \citep{dedecker2002necessary, Bulinski2017} in the proof of Theorem \ref{thm-asy-mu-hat-MDP}. This technique differs from that used in the proof of Theorem \ref{thm-asy-theta-hat-TL}.

\begin{theorem}
	\label{thm-asy-mu-hat-MDP}
	Under Assumptions \ref{ass-uniqueness}, \ref{assump-bounded-derivative-missing}, \ref{assump-covariate-distribution-missing} and model \eqref{DRM},
	we have
	$
	\sqrt{N}\left(\hat{\mu} - \mu\right)\stackrel{d}{\longrightarrow}\mathcal{N}(0,\tilde{\sigma}^{2})
	$
	as $N\rightarrow\infty$, where    $\tilde{\sigma}^{2} = \var\left(g(\bX, Y)\right)$.
\end{theorem}

Unlike the covariate shift case, 
we surprisingly find that  the  proposed W-estimator for $\mu$ 
is almost always asymptotically  super efficient  in the missing data setting.  
Let $\pi(\bx) = \mathbb{P}(D = 1\mid \bX = \bx)$.
\begin{corollary}
    \label{coro-super-efficient-MDP}
   Suppose that the assumptions in Theorem \ref{thm-asy-mu-hat-MDP} are satisfied.  If $\mathbb{P}(\pi(\bX) \neq 1)>0$ and $\mathbb{P}\left(g(\bX, Y)\neq g_{1}(\bX)\right)>0$, then the proposed W-estimator $\hat{\mu}$ is strictly more efficient than
	a semiparametric efficient estimator for $\mu$, or  is asymptotically  super efficient.
\end{corollary}

It is important to note that, given datasets $\{D_{i}, i\in[N]\}$, the  missing data framework can be reformulated as the covariate shift problem we previously addressed. This is achieved by designating 
$\{\bX_{i}:i\in \mathcal{N}_{0}\}$ 
 as the target sample and 
$\{(\bX_{i}, Y_{i}):i\in \mathcal{N}_{1}\}$
 as the source sample. Consequently, akin to Proposition \ref{prop-not-ral-TL}, we can demonstrate that 
$\hat{\mu}$ is not an RAL estimator, thereby substantiating the conclusion presented in Corollary \ref{coro-super-efficient-MDP}.

Furthermore, we  extend the enhanced W-estimation methods to the missing data setting, which is summarized in  Algorithm \ref{algorithm-data-splitting-missing}. Similar to Theorem \ref{thm-asymptotic-enhanced}, we can prove that the asymptotic distribution of $\hat{\mu}_{\rm en}$ is again the same as that of $\hat{\mu}$ regardless of  the regression method used.

\begin{algorithm}[H]
\label{algorithm-data-splitting-missing}
  \SetAlgoLined
  \KwData{$\{\bX_{i}:i\in \mathcal{N}_{0}\}, \{(\bX_{i}, Y_{i}):i\in \mathcal{N}_{1}\}$, a regression algorithm $\mathcal{A}$}
  \KwResult{$\hat{\mu}_{\rm en} $}
  
Split each of the sets $\mathcal{N}_{1}$ and $\mathcal{N}_{0}$ into two halves, denoted as $\mathcal{I}_1^{(1)}$, $\mathcal{I}_2^{(1)}$, $\mathcal{I}_1^{(0)}$, and $\mathcal{I}_2^{(0)}$, with corresponding sample sizes $n_1$, $n_2$, $m_1$, and $m_2$, respectively. 
 
Fit a regression model $\hat{\zeta}^{(i)}$   of $g(\bX, Y)$ on $\bX$ with method   $\mathcal{A}$  
based on  data  $\{( \bX_{j},  g(\bX_j, Y_j)): j\in \mathcal{I}_{i}^{(1)}\}$ for $i=1,2$.

Calculate  
$\hat{q}_{l}^{(i)} =  (1/m_{i}) \sum_{j\in\mathcal{I}_{i}^{(0)}}  \prod_{k\in\mathcal{I}_{i}^{(1)}}I(\tilde e_{lj}\leq \tilde e_{kj})$,   $l\in\mathcal{I}_{i}^{(1)}$,   $i=1, 2$. 

Let  $N_{1} = n_{1}+m_{1}$ and $N_{2} = n_{2}+m_{2}$ and calculate the enhanced W-estimate for $\mu$ in each sample, 
\bas
\hat{\mu}^{(1)}_{\rm en} &=& \frac{1}{N_{1}}\sum\limits_{j\in\mathcal{I}_{1}^{(1)}}g(\bX_{j}, Y_{j})+\frac{1}{N_{1}}\sum\limits_{j\in\mathcal{I}_{1}^{(0)}}\hat{\zeta}^{(2)}(\bX_{j})+\frac{m_{1}}{N_{1}}\sum\limits_{i\in\mathcal{I}_{1}^{(1)}}\hat{q}_{i}^{(1)}\{g(\bX_{i}, Y_{i}) - \hat{\zeta}^{(2)}(\bX_{i})\},\\
\hat{\mu}^{(2)}_{\rm en} &=& \frac{1}{N_{2}}\sum\limits_{j\in\mathcal{I}_{2}^{(1)}}g(\bX_{j}, Y_{j})+\frac{1}{N_{2}}\sum\limits_{j\in\mathcal{I}_{2}^{(0)}}\hat{\zeta}^{(1)}(\bX_{j})+\frac{m_{2}}{N_{2}}\sum\limits_{i\in\mathcal{I}_{2}^{(1)}}\hat{q}_{i}^{(2)}\{g(\bX_{i}, Y_{i}) - \hat{\zeta}^{(1)}(\bX_{i})\}. 
\eas

Return the data-splitting based W-estimate
$
\hat{\mu}_{\rm en} =  (N_{1}/N)\hat{\mu}^{(1)}_{\rm en}+ (N_{2}/N)\hat{\mu}^{(2)}_{\rm en}.
$

\caption{Data-splitting based W-estimator for missing data problems} \end{algorithm}

\section{Simulation studies}
\label{sec-simu}
In this section, we perform simulation studies to empirically validate the main theoretical findings and compare the performance of the proposed W-estimator against competing methods under both covariate shift and MAR  settings.

\subsection{Empirical verification of theoretical results}

We first examine  the consistency and asymptotic normality of the proposed W-estimator $\hat{\theta}$, as established in Theorem~\ref{thm-asy-theta-hat-TL}, and then investigate the consistency of the variance estimator  $\hat{\sigma}^2$ presented in Proposition~\ref{prop-var-est}.

We consider a covariate shift setting with two  covariates, $\bX=(X_{(1)},X_{(2)})$.
Under the source distribution, $X_{(1)}$ and $X_{(2)}$ are independent,   each following a $\mathrm{Beta}(2,3)$ distribution.
Under the target distribution, the covariates remain independent, with $X_{(1)},X_{(2)} \sim \mathrm{Beta}(3,4)$.
Conditional on $\bX=\bx$, the response variable satisfies  
\[
Y | \bX=\bx \sim N(f(\bx), 0.1^2)\quad\text{with}\quad f(\bx)=\log(x_{(1)}+x_{(2)}).
\]
We set the source sample size  $n$ equal to the target sample size $m$, and consider values of $m$ in the set  $\{50, 100, 500, 1000, 3000, 5000, 7000, 9000 \}$. For each $m$, the experiment is repeated independently $R = 3000$ times. In each replication, we compute the Vanilla W‑estimator  $\hat{\theta}$ as defined in Equation~\eqref{eq:theta-hat-w-vanilla} in Section~\ref{sec:minimum-w-dist-estimate}.

\begin{figure}[!ht]
\centering
\includegraphics[width=0.45\linewidth]{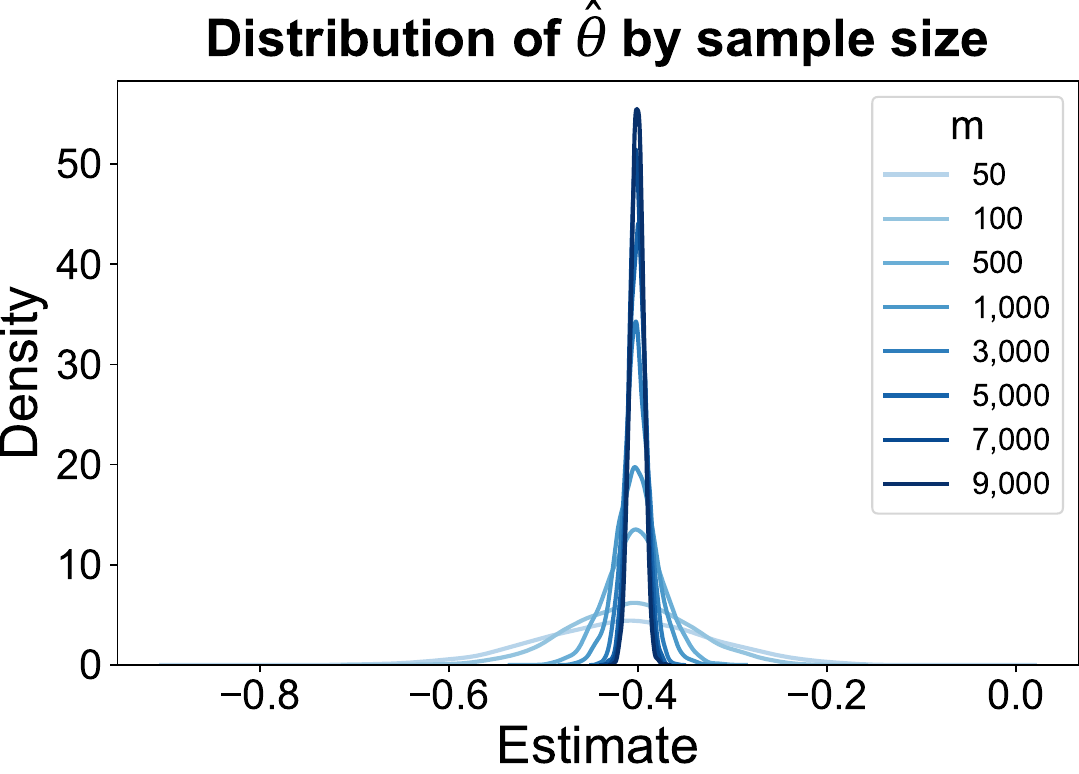}
\includegraphics[width=0.45\linewidth]{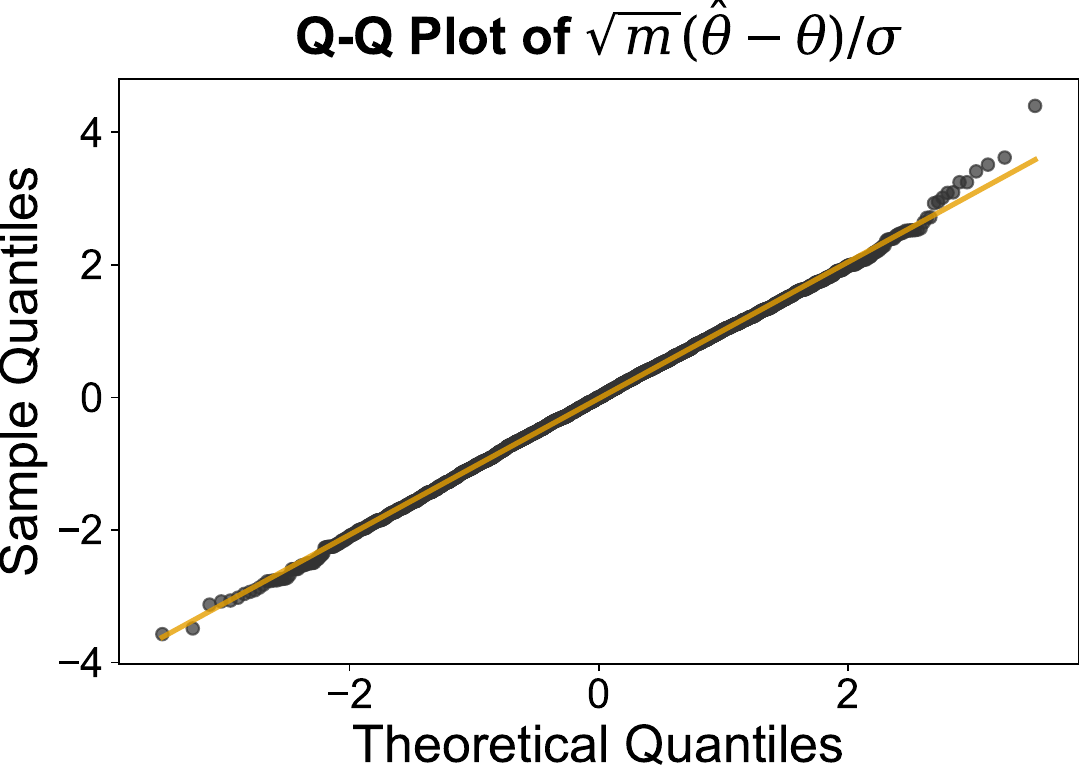}
\caption{Empirical verification of Theorem~\ref{thm-asy-theta-hat-TL}.}
\label{fig:theorem2_verification}
\end{figure}

{\it Consistency}. 
Figure~\ref{fig:theorem2_verification} (left) presents the kernel density estimates (KDEs) of  $\hat{\theta}$  across the $R$ Monte Carlo replications for various sample sizes $m$. The KDEs are constructed with a Gaussian kernel, using Scott's rule of thumb to select the bandwidth. As $m$ grows, the distribution of  $\hat{\theta}$   concentrates progressively around the true parameter value ($-0.4008$), offering clear empirical support for its consistency.

{\it Asymptotic normality}.
To further evaluate the asymptotic normality of $\hat{\theta}$, we compute the  asymptotic standard deviation $\sigma$ via a high‑precision Monte Carlo approximation. The estimator is standardized as $\sqrt{m}(\hat{\theta} - \theta)/\sigma$. Figure~\ref{fig:theorem2_verification} (right) presents the Q--Q plot of the standardized W‑estimator against a standard normal distribution for $m = 9000$. The points align closely with the $45^\circ$ reference line, with only minor deviations in the extreme tail; this is consistent with the asymptotic normality behavior described in Theorem~\ref{thm-asy-theta-hat-TL}. Overall, these results provide strong empirical support for the asymptotic normality conclusion.

\begin{figure}[!ht]
\centering
\includegraphics[width=0.45\linewidth]{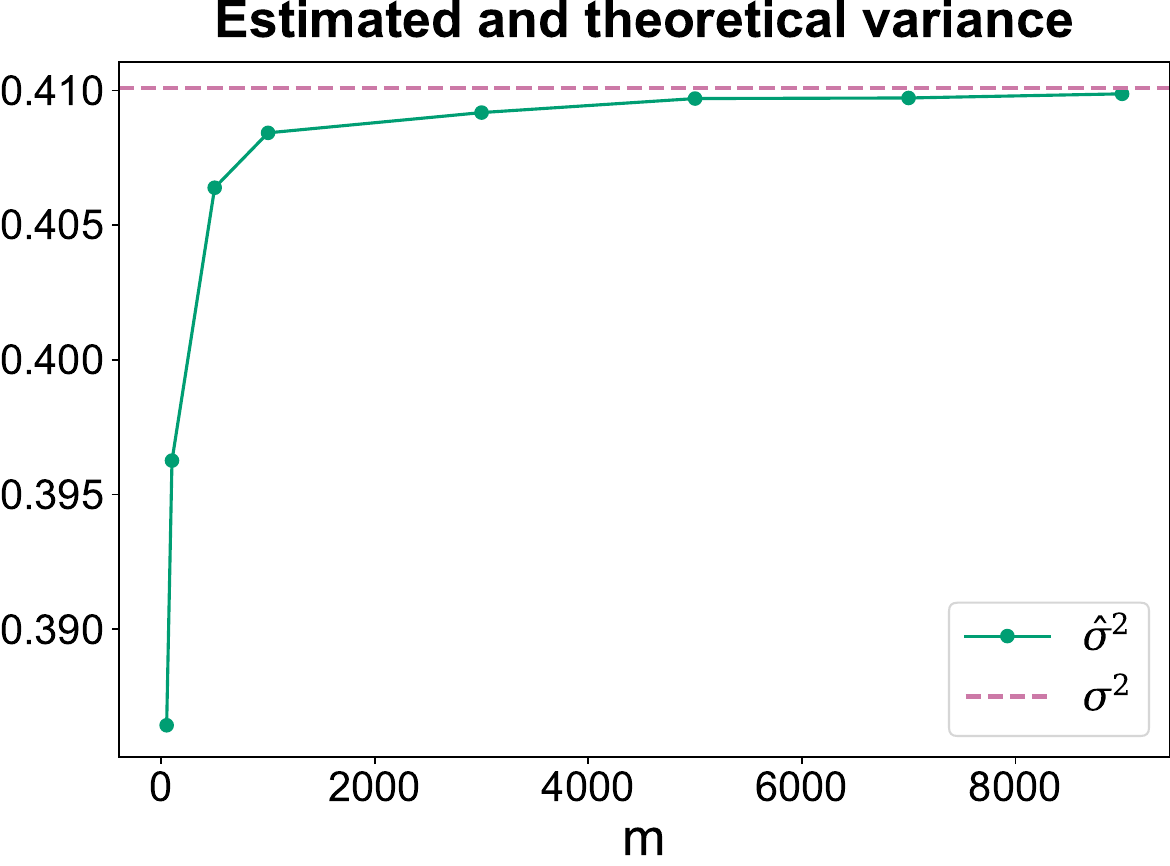}
\includegraphics[width=0.45\linewidth]{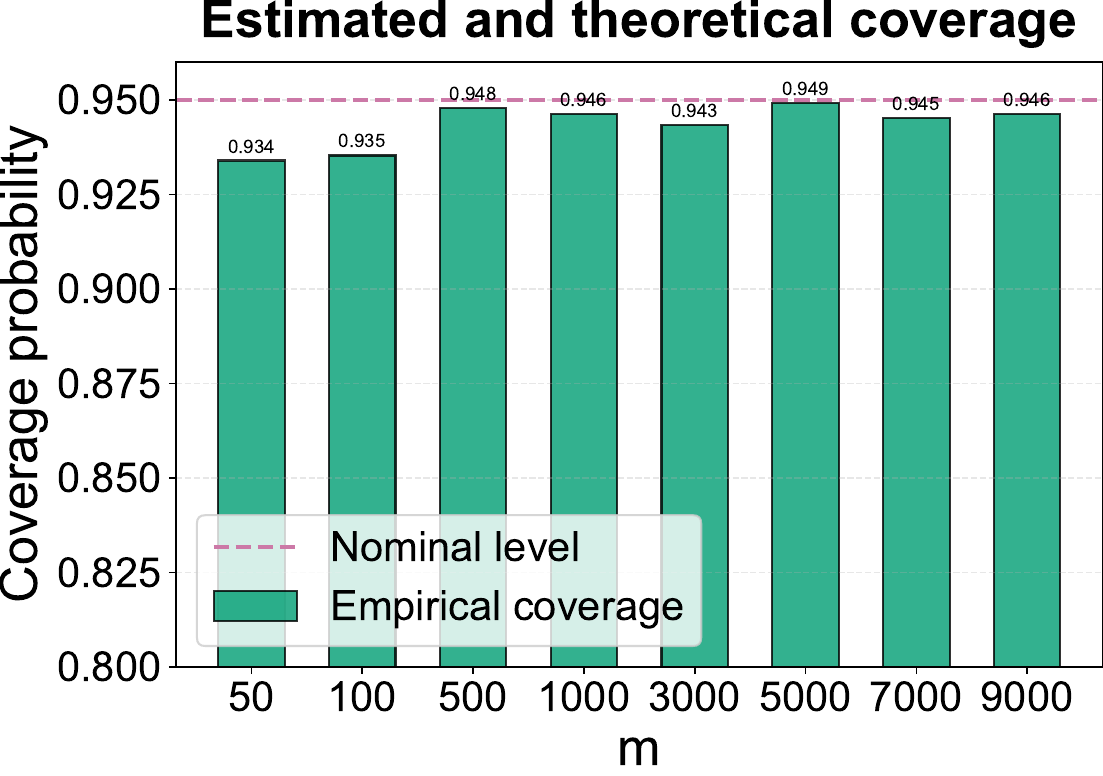}
\caption{Empirical verification of Proposition~\ref{prop-var-est}.}
\label{fig:prop2_verification}
\end{figure}

We next  examine the consistency of the variance estimator  $\hat{\sigma}^2$ presented in Proposition~\ref{prop-var-est}. 
We employ the same simulation setup described earlier and compute  $\hat{\sigma}^2$  in each Monte Carlo replication using Equation~\eqref{eq:hat-var}. Figure~\ref{fig:prop2_verification} (left) displays the average of  $\hat{\sigma}^2$ across $R$ replications as a function of $m$, with the true variance $\sigma^2$ ($0.4101$)  represented by a dashed horizontal line. As the sample size grows, the estimated variance converges to the true value, providing empirical confirmation of the consistency of  $\hat{\sigma}^2$, in alignment with Proposition~\ref{prop-var-est}. 
By combining Theorem~\ref{thm-asy-theta-hat-TL} and Proposition~\ref{prop-var-est}, we construct Wald‑type confidence intervals
$
\hat\theta \pm 1.96 \, \hat\sigma/\sqrt{m}
$
for \(\theta\) at the 95\% level and evaluate their empirical coverage.
Figure~\ref{fig:prop2_verification} (right) shows that the empirical coverage closely matches the nominal level, offering clear empirical support for the theoretical results.

\subsection{Covariate shift}

We now evaluate the empirical performance of the proposed method under covariate shift, employing scenarios designed to cover a range of practically relevant settings, with particular emphasis on efficiency and robustness to model misspecification.

We again consider a bivariate covariate vector \(\bX = (X_{(1)}, X_{(2)})\). The source distribution of \(\bX\) is set to \(N((0,0)^{\top}, 0.5^2 I_2)\), while the target distribution is \(N((0.2,0)^{\top}, 0.3^2 I_2)\). Given \(\bX = \bx\), the response is generated as \(Y \mid \bX = \bx \sim N(f(\bx), 0.1^2)\), where \(f(\bx)\) is chosen from the following six functions:
\begin{itemize}
    \item [(i)] \(x_{(1)}+x_{(2)}\) \quad \quad\quad\quad\quad~~~ (ii) \(\log\big((x_{(1)}+x_{(2)})^2\big)\)
    \item [(iii)] \(\sin(x_{(1)}+x_{(2)}) + 2 x_{(1)}^2\) \quad (iv) \(\cos((x_{(1)}+x_{(2)}) / 4) + \log(x_{(1)}^2+x_{(2)}^2)\)
    \item [(v)] \(2 + |x_{(1)}+x_{(2)}|\) \quad\quad\quad\quad  (vi) \(2 \sin(x_{(1)}+x_{(2)} + 2)\sin(x_{(1)} + 1) +\cos(x_{(1)}+x_{(2)})\)
\end{itemize}
These functions span a wide range of structural complexity: (i) is linear, (ii) is mildly nonlinear with a log-square transformation, (iii) combines smooth oscillation with a quadratic term (exhibiting high curvature), (iv) merges low-frequency oscillation with a radial logarithmic term (non-separable and varying in smoothness near the origin), (v) introduces non-differentiable, piecewise-linear behavior via the absolute value, and (vi) is highly oscillatory with multiplicative interactions that produce multiple modes. Collectively, they vary in smoothness, curvature, nonlinearity, and interaction complexity, offering a diverse set of challenges for regression.

We generate \(n\) paired labelled observations  $(\bX, Y)$ from the source distribution  and \(m\) unlabelled observations $\bX$ from the target distribution. In each experiment, we vary the source and target sample sizes \((n, m)\) over the set  
\( \{(100,100), (100,500), (500,500), (500,1000),  \\ 
(500,2000), (1000,1000)\} \).
We make inference about $\theta = \e_1 \{g(\bX, Y)\}$, and    
 consider two choices for  $g(\bX, Y)$:  (1) $g(\bX, Y)=Y$ so that  $\theta$ 
is the response mean of the target population, 
and (2) $g(\bX, Y)=X_{(1)}Y$, 
so that   $\theta $  is proportional to the correlation coefficient of $X_{(1)}$ and $Y$.

To assess the empirical performance, we compare our proposed estimator, the Vanilla W-estimator (W-V), and the Enhanced W-estimator (W-E) with several widely used methods for handling covariate shift.  
Let  $m(\bx) = \e\{ g(\bX,Y)|\bX=\bx \}$ and  $r(\bx) = dF_{1}(\bx)/dF_{0}(\bx)$. 
The competing estimators are described below: 
\begin{itemize}
\item   
\textbf{Na\"ive}:      $\hat\theta^{\text{naive}} = n^{-1}\sum_{i=1}^{n} g(\bX_{i},Y_{i})$,
which  ignores the distributional discrepancy and uses only the labeled source data.

\item  
\textbf{Pseudo-labeling (PL)} \citep[Section 4.3]{Yu2012607}: 
$\hat{\theta}^{\text{PL}} = m^{-1}\sum_{j=1}^{m}\hat{m}(\tilde\bX_{j})$, where $\hat{m}(\bx)$ is an estimator 
of  $m(\bx)  $   trained on the source data.

\item \textbf{Importance weighting (IW)}~\citep{Shimodaira2000Improving}: 
$\hat{\theta}^{\text{IW}} = n^{-1}\sum_{i=1}^{n}\hat{r}(\bX_{i})g(\bX_i,Y_{i})$,
where   $\hat{r}(\bx)$ is an estimator of  the density ratio $r(\bx)  $. 
We estimate $r(\bx)$ using two standard techniques: KL-divergence minimization~\citep{sugiyama2007direct} and discriminative classification-based estimation~\citep{bickel2009discriminative}.
The resulting estimators for $\theta$ are named  as IW-KL and IW-DC, respectively.

\item \textbf{Double robust (DR)}~\citep{li2020robust}. 
This estimator combines importance weighting and pseudo-labeling via sample splitting.
The source sample is divided into two equal halves: the first half, together with the target covariates, is used to estimate the density ratio $\hat{r}$, and the second half is used to fit a regression model
$\hat{m}$ for $m(\bx)  $.
The final estimator takes the form: 
\[
\hat{\theta}^{\text{DR}} = \frac{1}{\lfloor n/2 \rfloor}\sum\limits_{i=1}^{\lfloor n/2\rfloor}\hat{r}\{ 
\bX_{i})\{ g(\bX_i,Y_{i}) - \hat{m}(\bX_{i}) \} +\frac{1}{m}\sum\limits_{j=1}^{m}\hat{m}(\tilde\bX_{j}),
\]
where  $\hat{r}$ is obtained by  the discriminative classification method~\citep{bickel2009discriminative}.
\end{itemize} 
We repeat the data-generating process $R = 500$ times. 
For the $j$th replication, we generate samples from the source and target distributions and compute the corresponding estimate $\hat{\theta}_{j}$. 
The mean squared error (MSE) of an estimator is then approximated by
\(
    (1/R)\sum_{j=1}^{R} (\hat{\theta}_{j} - \theta^{*})^{2},
\) 
where $\theta^{*}$ denotes the true parameter value. 
Lower MSE values indicate better estimator performance.

{\it Efficiency comparison results}
We begin by comparing the efficiency of the different estimators for the target response mean $\theta = \mathbb{E}_{1}(Y)$.
For our proposed estimator, as well as the baseline methods such as PL and DR, we require a regression model to estimate the conditional mean. Given the diverse and potentially complex shapes of the regression functions under consideration, we employ XGBoost \citep{Chen-Guestrin-2016}, which typically delivers strong predictive performance across a wide range of regression tasks and is computationally efficient. The hyperparameters  of XGBoost in all methods are selected  through cross-validation. 
\begin{figure}[!ht]
\centering
\includegraphics[width=0.4\linewidth]{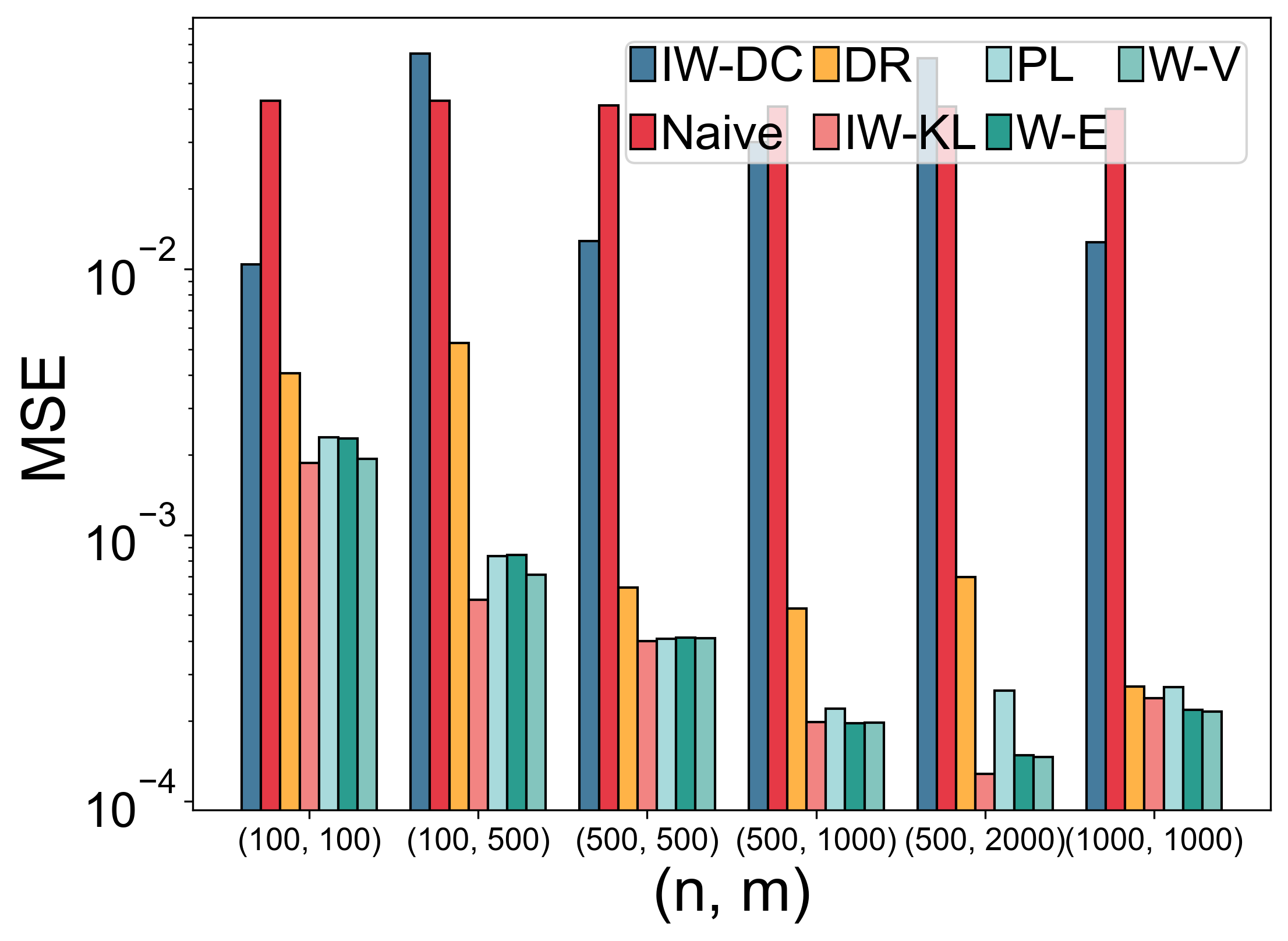}
\includegraphics[width=0.4\linewidth]{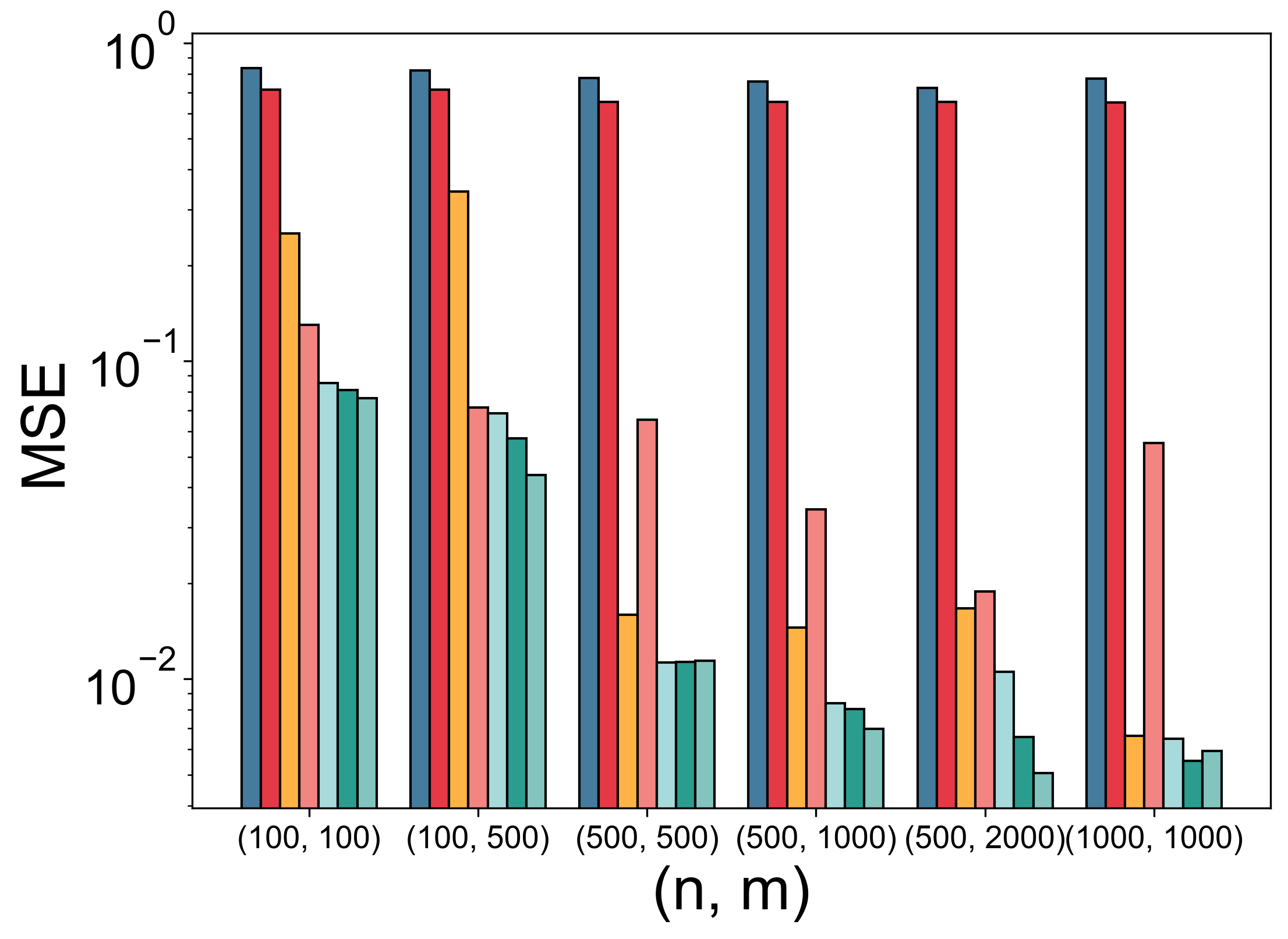}\\
\includegraphics[width=0.4\linewidth]{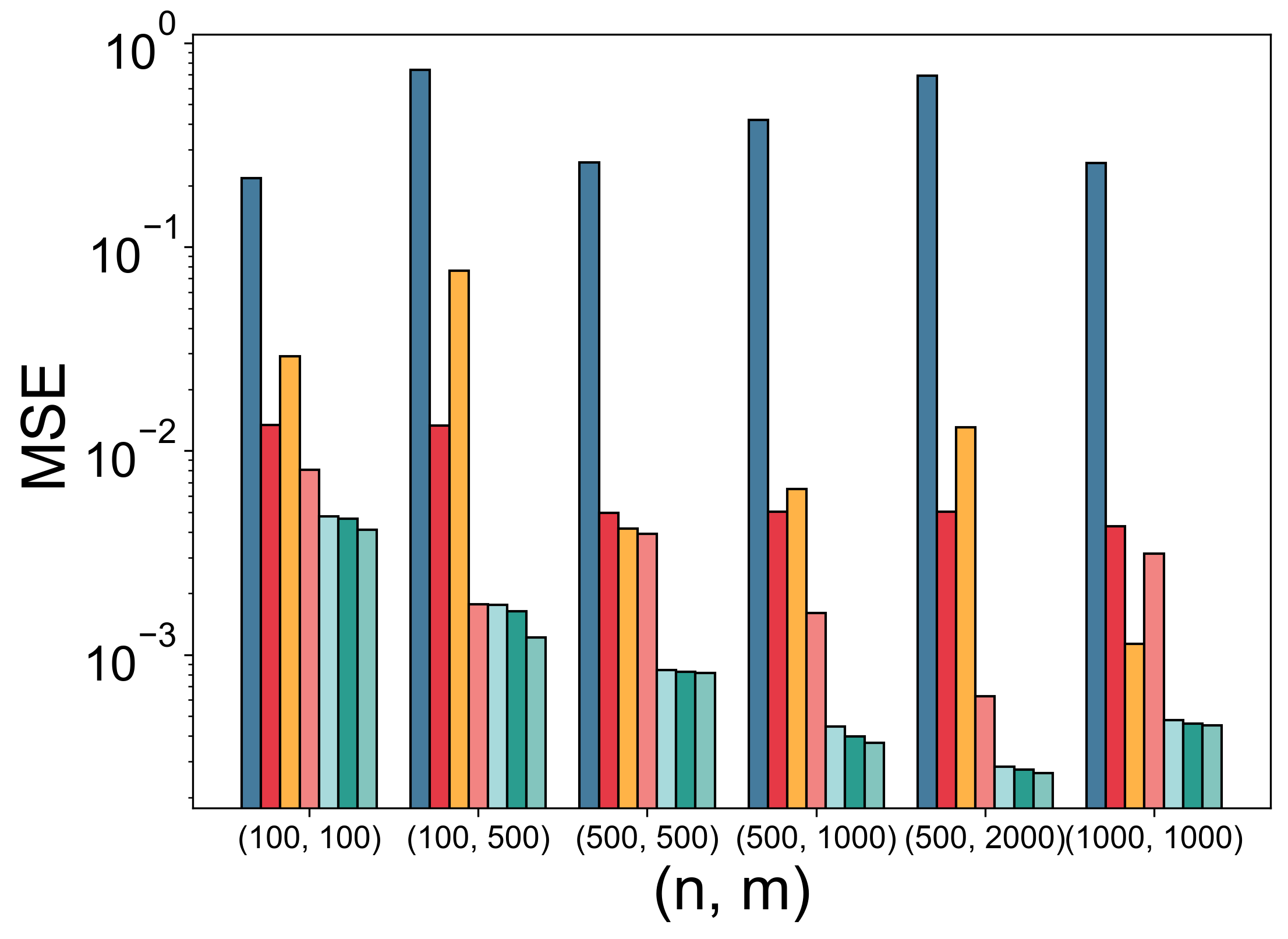}
\includegraphics[width=0.4\linewidth]{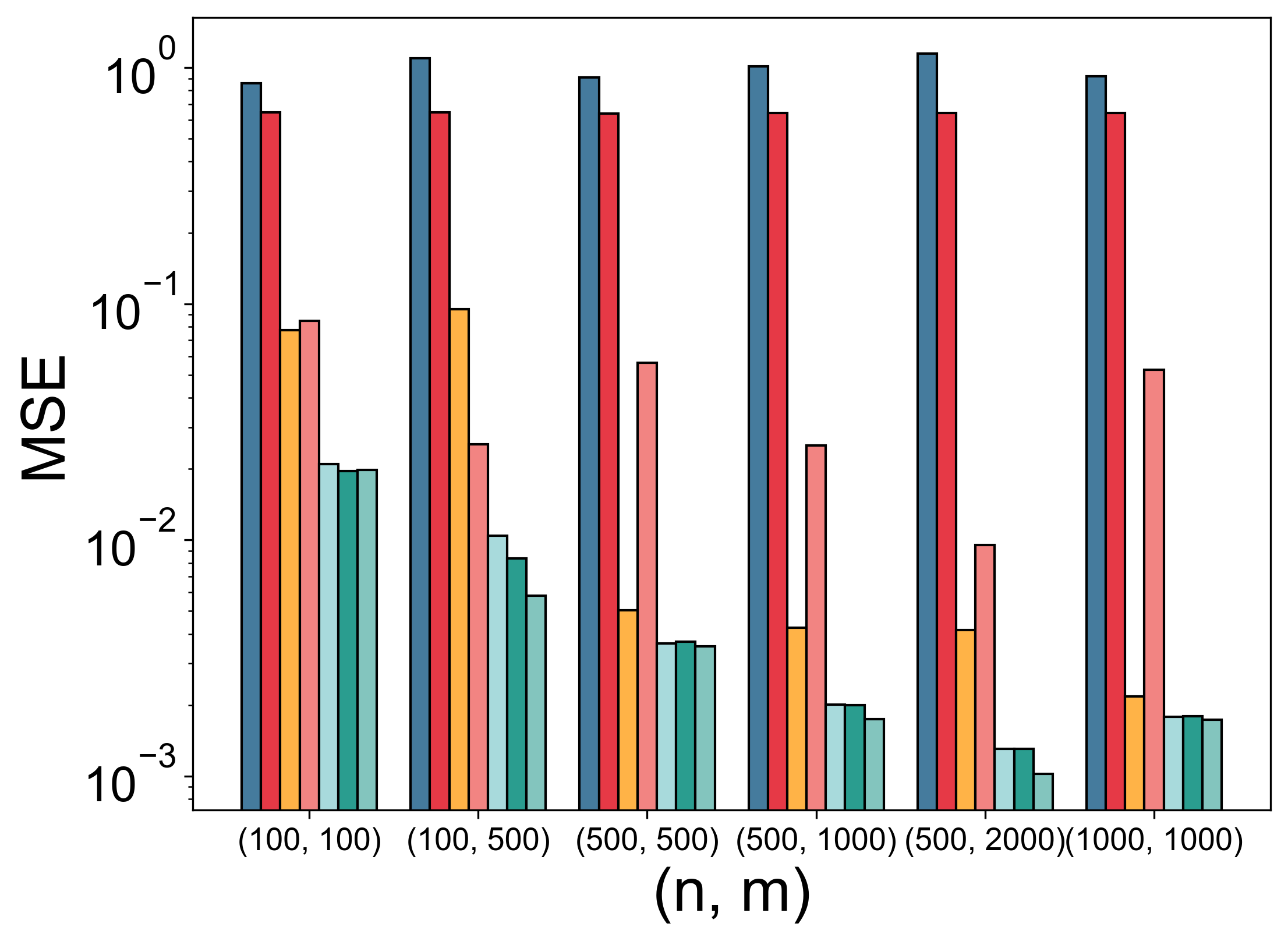}\\
\includegraphics[width=0.4\linewidth]{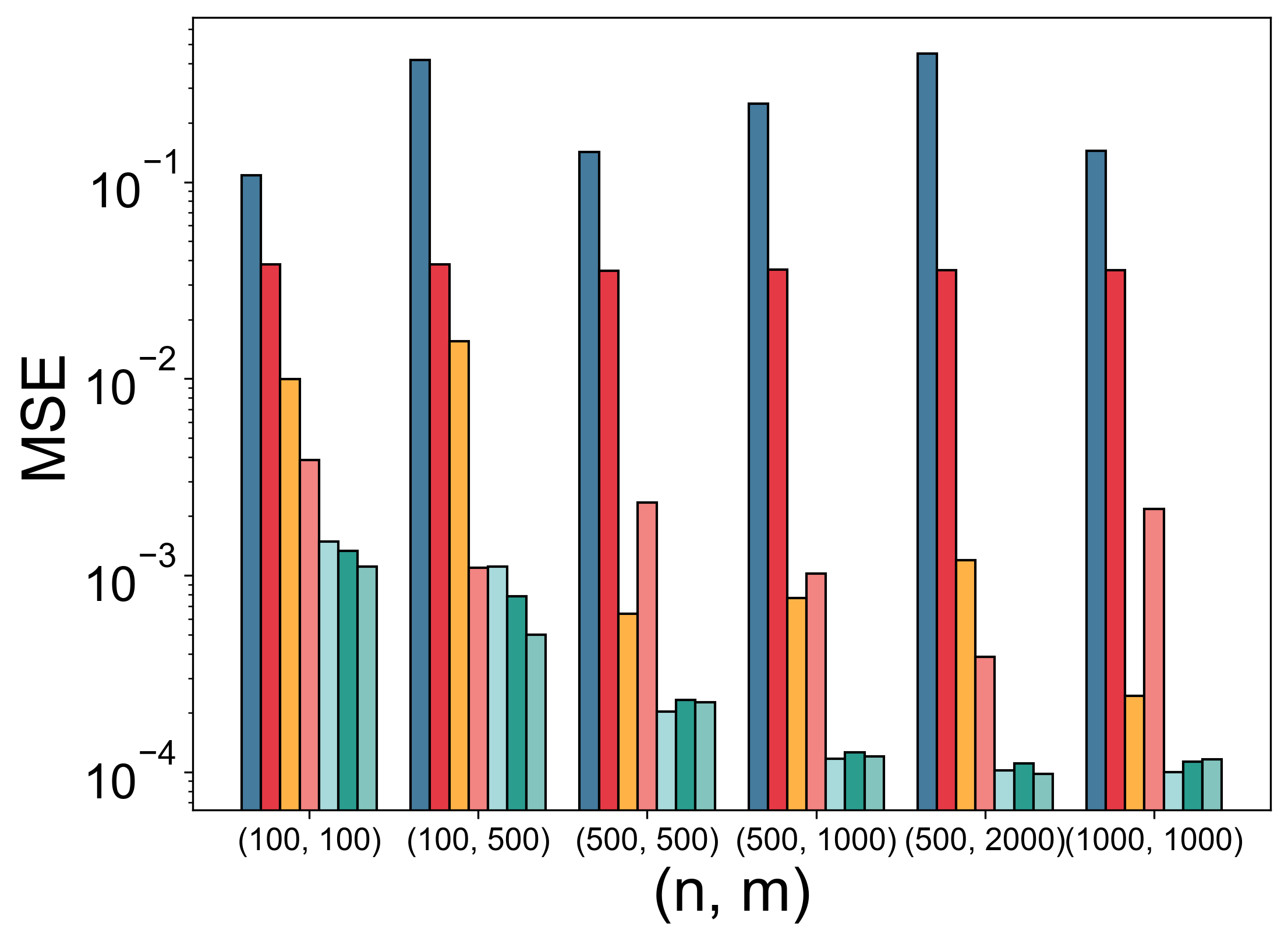}
\includegraphics[width=0.4\linewidth]{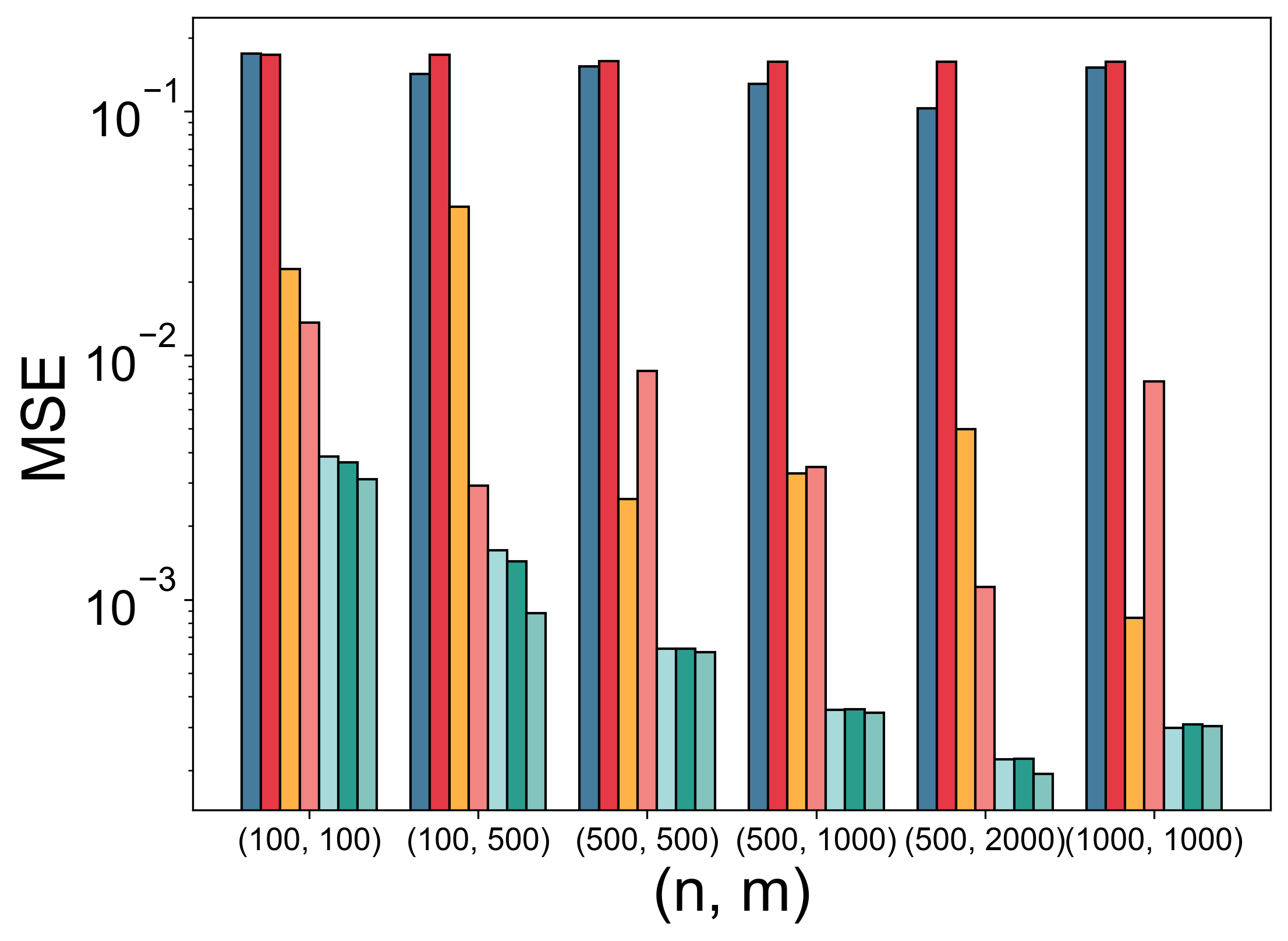}
\caption{MSE of various estimators under regression functions (i)--(vi) from left to right, top to bottom when \emph{XGBoost} is used for regression. Within each panel, results are shown for different combinations of source sample size $n$ and target sample size $m$.}
\label{fig:efficiency_covariate_shift}
\end{figure}
In this experiment, the sample sizes \(n\) and \(m\) are chosen to satisfy the conditions in Corollary~\ref{super-efficiency-TL}. The results are summarized in Figure~\ref{fig:efficiency_covariate_shift}. Overall, the proposed W‑E and W‑V estimators consistently outperform the baseline methods in nearly all scenarios. The sole exception arises for simple regression functions, such as the linear case (i), where IW‑KL occasionally achieves marginally better performance. These observations align well with the theoretical analysis provided in Corollary~\ref{super-efficiency-TL}.

\begin{figure}[!ht]
\centering
\includegraphics[width=0.4\linewidth]{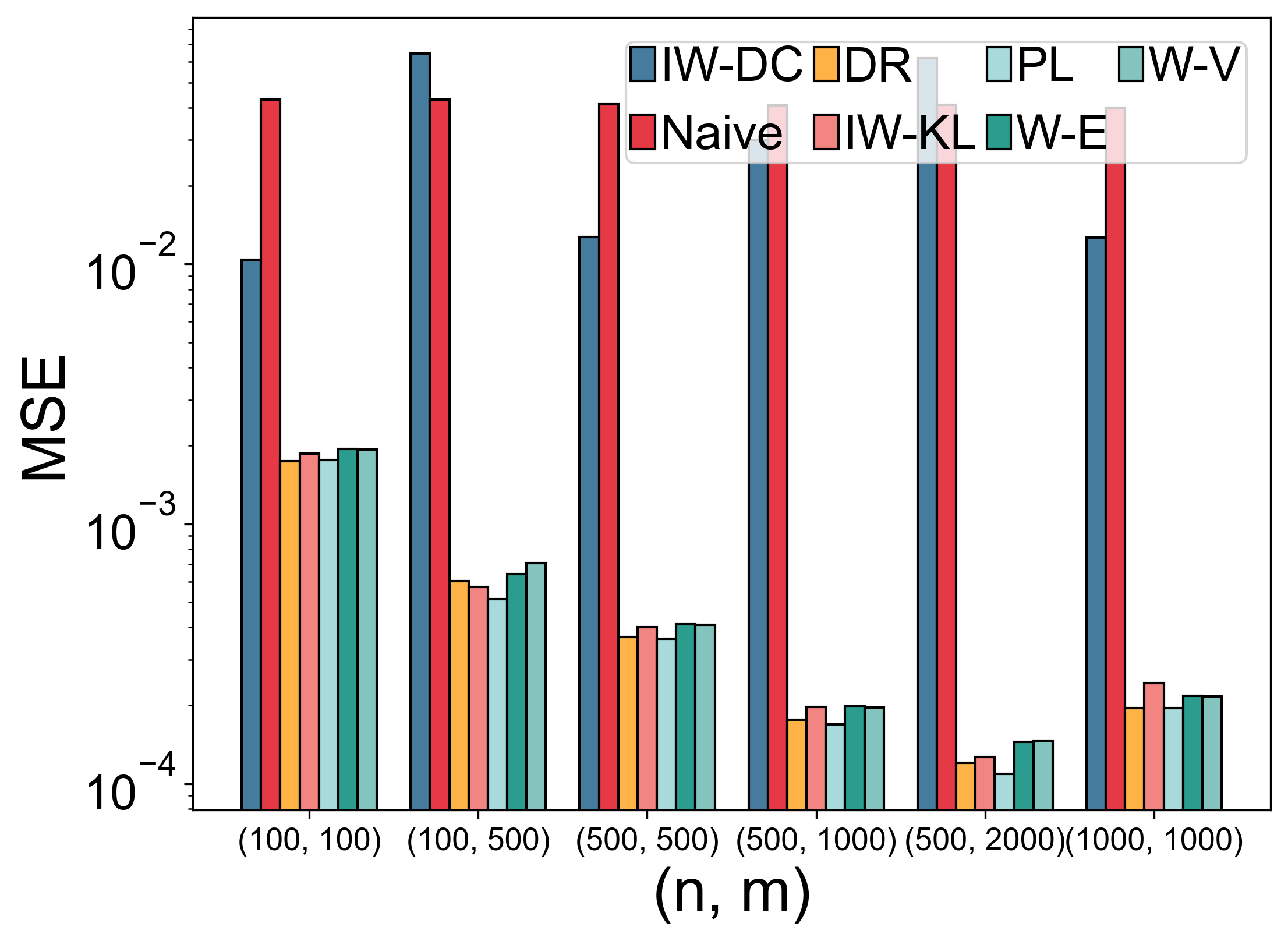}
\includegraphics[width=0.4\linewidth]{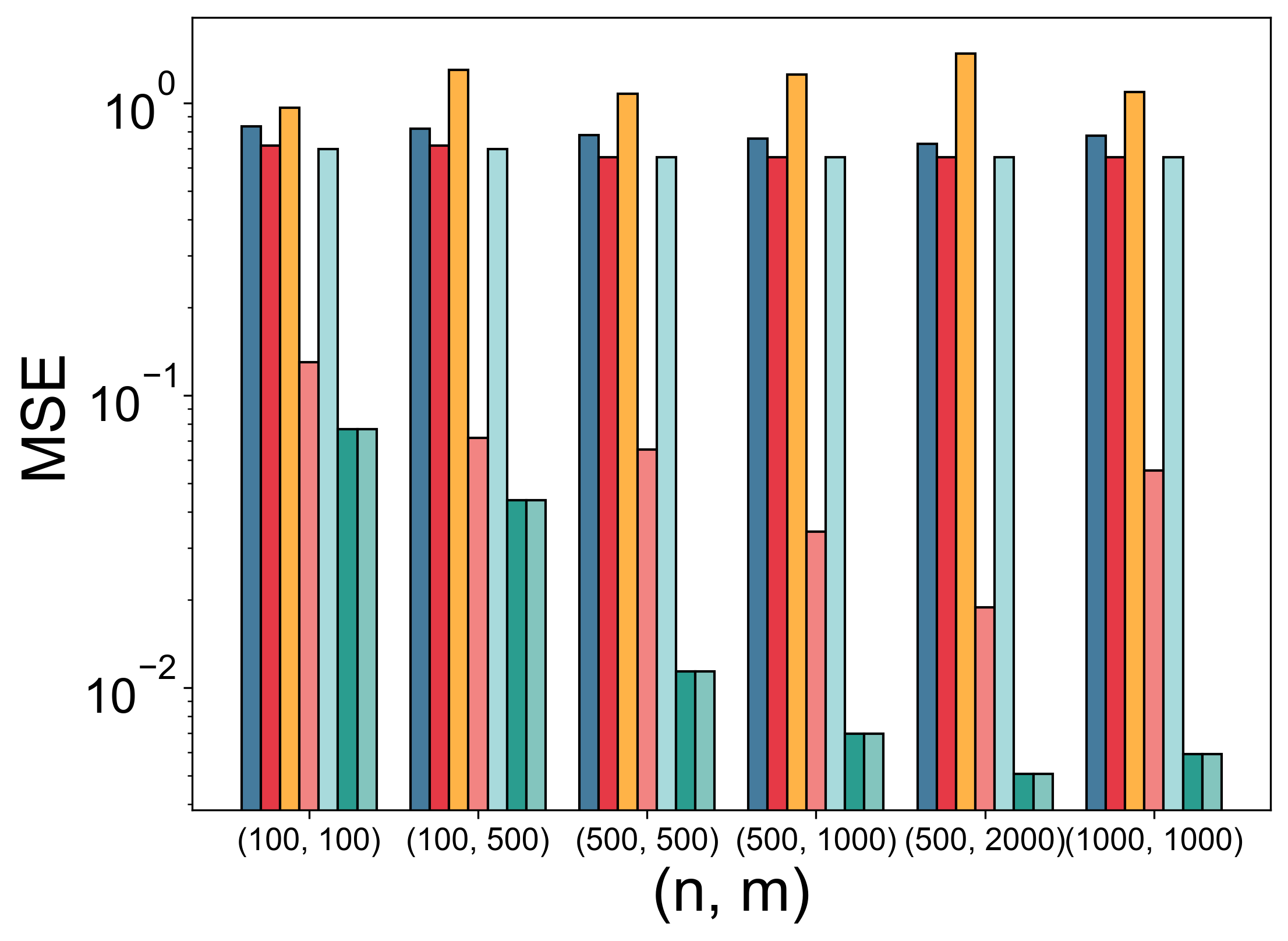}\\
\includegraphics[width=0.4\linewidth]{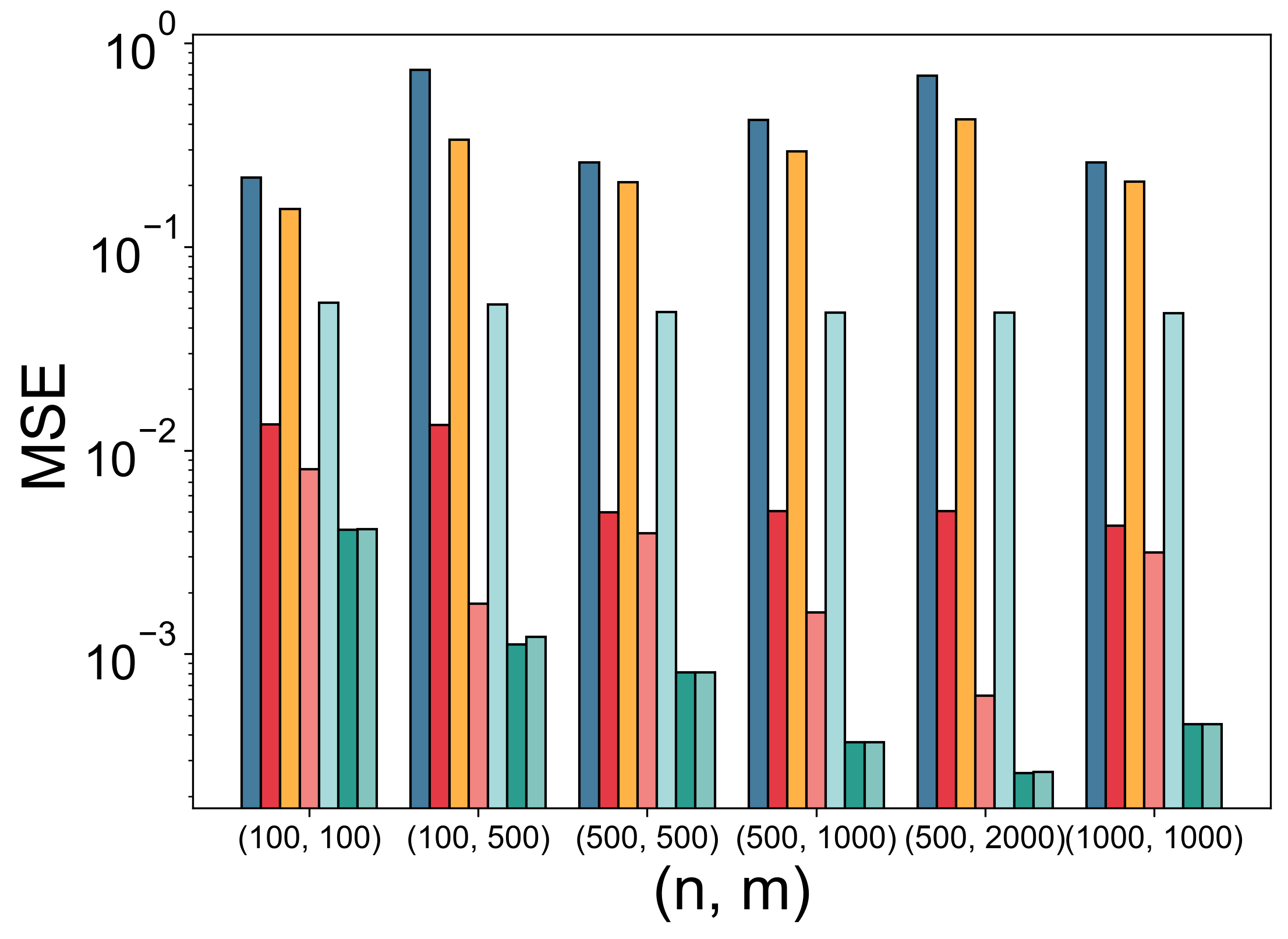}
\includegraphics[width=0.4\linewidth]{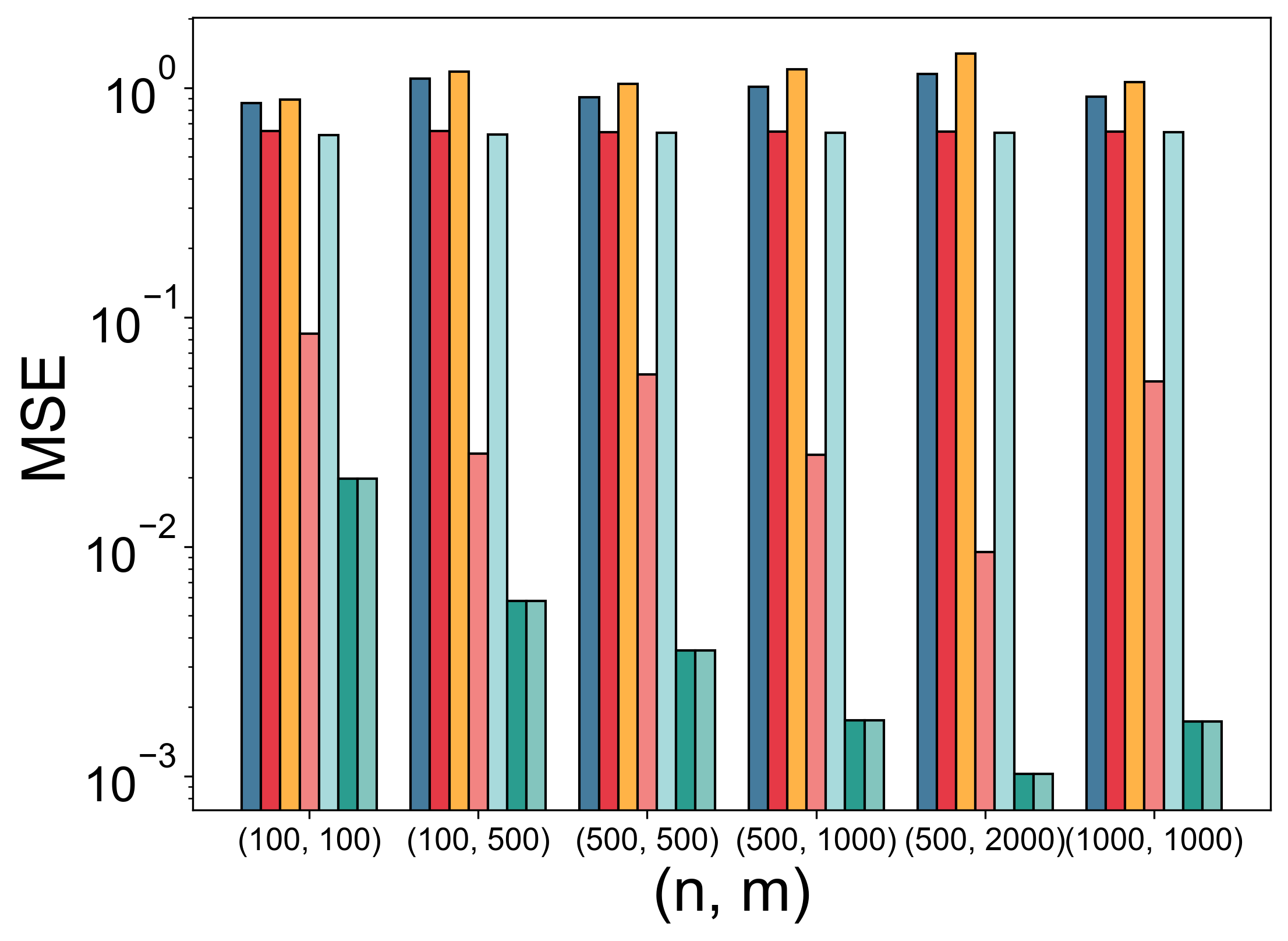}\\
\includegraphics[width=0.4\linewidth]{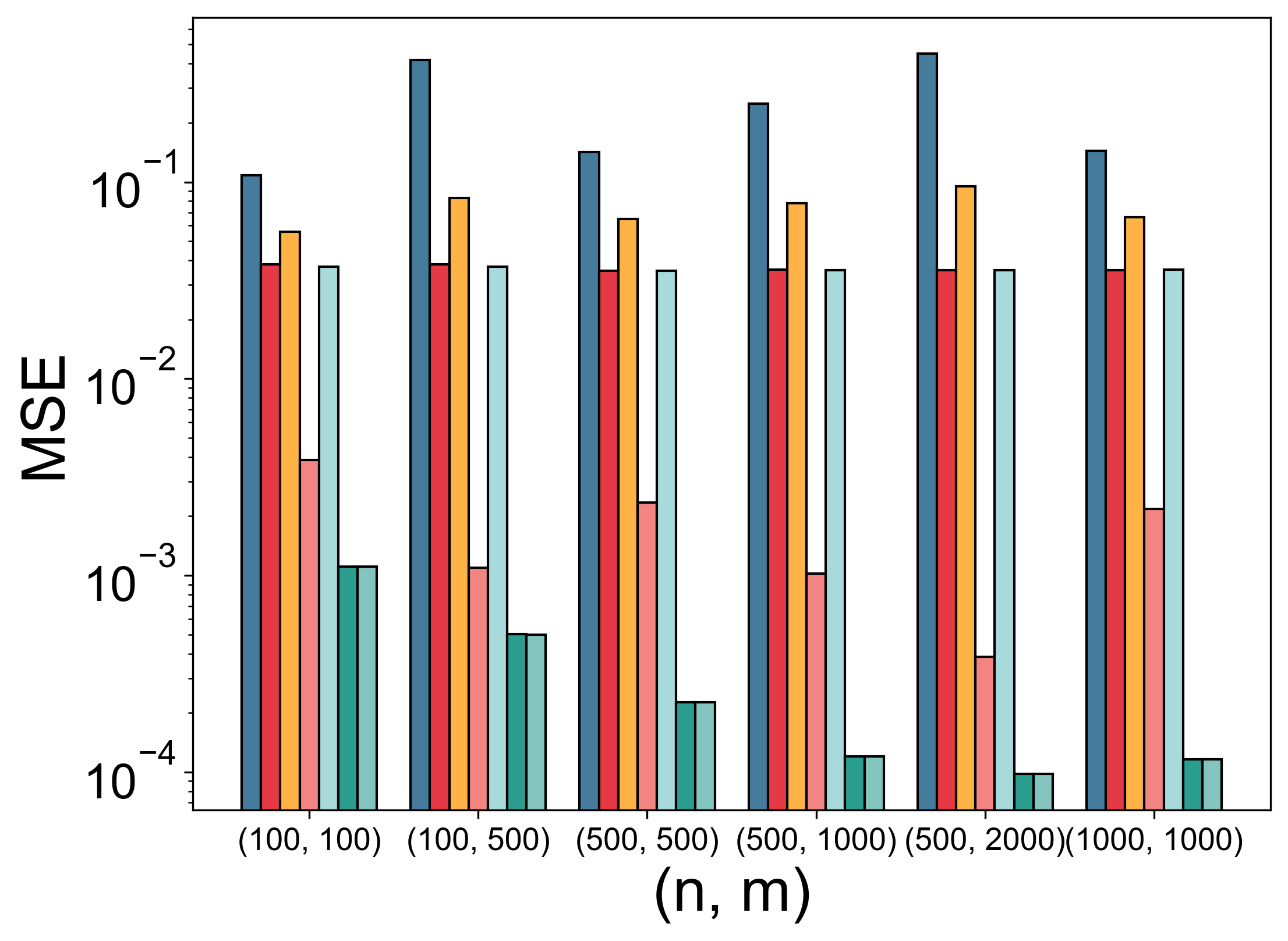}
\includegraphics[width=0.4\linewidth]{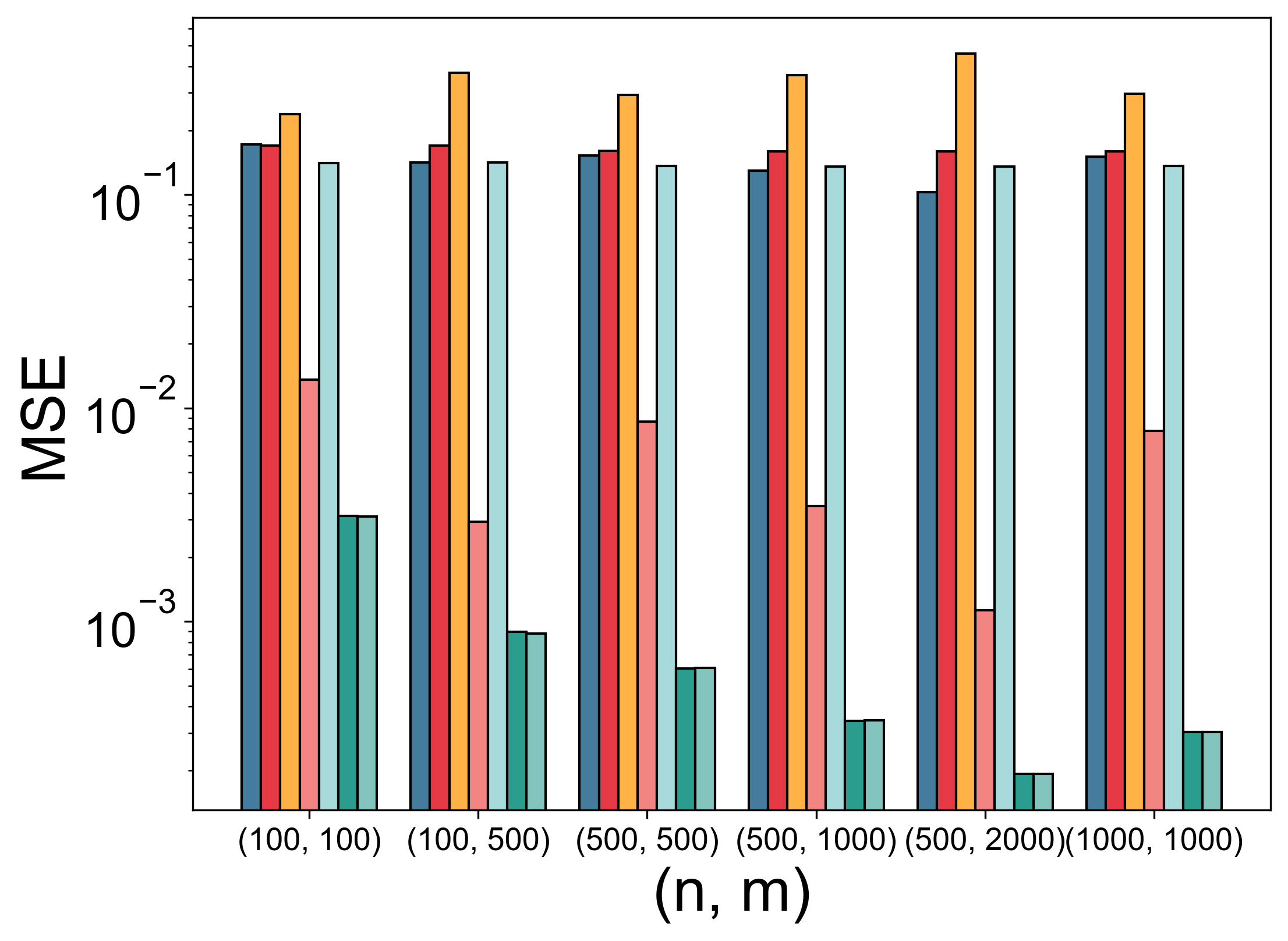}
\caption{MSE of various estimators under regression functions (i)--(vi) from left to right, top to bottom when \emph{linear model} is used for regression. Within each panel, results are shown for different combinations of source sample size $n$ and target sample size $m$.}
\label{fig:robustness_covariate_shift}
\end{figure}

{\it Robustness comparison results}
We next compare the robustness of different estimators for the correlation $\theta = \e_{1}\{X_{(1)}Y\}$ when the regression model $m(\bx)$ is mis-specified.
In all cases, the true regression function is nonlinear in the covariates, yet we fit a linear regression model.   
The results are presented in Figure~\ref{fig:robustness_covariate_shift}.
When the regression model is mis-specified, the advantage of our proposed methods becomes more pronounced. 
These results indicate that the proposed W-E and W-V estimators not only achieve super-efficiency under correct specification but also exhibit robustness to model misspecification.

\subsection{Missing data}
We next study the empirical performance of the proposed estimators in a missing data setting. 
The MAR setting can be viewed as a special case of covariate shift, in which selection induces different marginal distributions for the covariates across observed and unobserved samples while the conditional outcome distribution $Y\mid \bX$ remains invariant. 
Since this structural form has already been extensively studied in the covariate shift experiments in the previous subsection, we do not repeat those experimental settings here. 
Instead, we consider a complementary regime that is specific to missing data problems: all baseline approaches are equipped with correctly specified base learners for both the outcome regression and the missingness mechanism. 
This allows us to isolate and evaluate the efficiency behavior of the proposed method in a canonical MAR setting and to assess its performance under conditions where standard estimators are expected to be well behaved.

We generate $\bX = (X_{(1)},  X_{(2)})^\top$ from a bivariate normal distribution $ N\big((0,0)^\top, 0.5^2 I_2\big)$.  
Given $\bX$, we generate the outcome $Y$ from a linear model   
$Y = 1 + X_{(1)} + X_{(2)} + \varepsilon$  with  $\varepsilon \sim N(0, 0.1^2)$.
The outcome $Y$ is subject to missingness and the nonmissingness indicator $D$ satisfies 
 a MAR mechanism:
\[
\mathbb{P}(D=1 \mid \bX, Y) = \mathbb{P}(D=1 \mid \bX) = \pi(\bX),
\]
where the propensity score follows a logistic regression model
\[
\text{logit}\{\pi(\bX)\} =  0.5 +  0.1\times X_{(1)} + 1.2\times X_{(2)}. 
\]

The target parameter of interest is the marginal mean of the outcome, $\theta = \e(Y)$.
The outcome regression model $ m(\bx)$ is specified as a linear model (the fitted model is denoted as
$\hat{m}(\bX)$), while the propensity score model is a logistic regression with a linear predictor in the covariates.
Sample sizes are set as $N \in \{1000, 3000, 5000, 7000, 9000\}$, and for each setting, the experiment is repeated independently $R = 500$ times.

We compare the proposed estimators against the following classical and modern methods for handling missing outcomes:
(1) a pseudo-labeling (PL) estimator,
$
\hat{\theta}^{\text{PL}} = N^{-1} \sum_{i=1}^N \{ D_i Y_i + (1 - D_i) \hat{m}(\bX_i) \},
$
(2) the inverse probability weighting (IPW) estimator,
(3) the augmented inverse probability weighting (AIPW) estimator,
and
(4) the double machine learning (DML) estimator \citep{Chernozhukov2018} with two-fold cross-fitting.

\begin{figure}[!ht]
\centering
\includegraphics[width=0.6\linewidth]{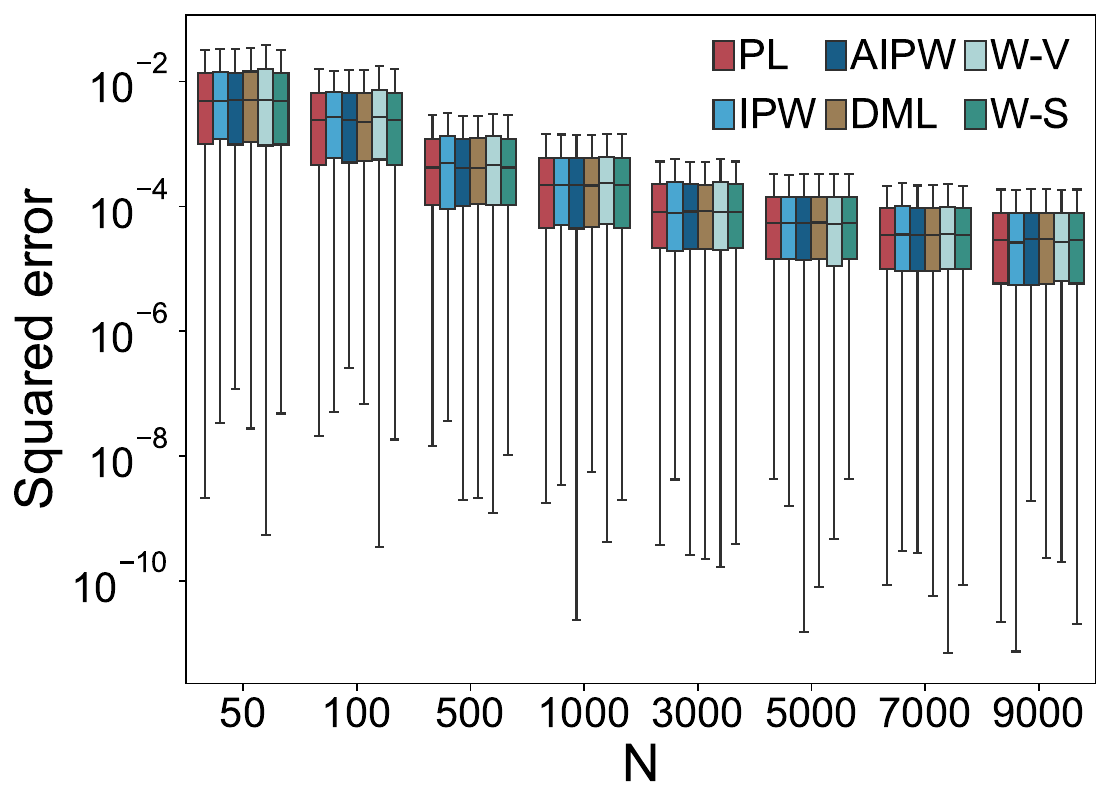}
\caption{Boxplots of the SE $(\hat\theta_r - \theta^*)^2$ for various estimators over $R=500$ repetitions under the MAR setting, across different sample sizes, with both the outcome regression and propensity score models correctly specified.}
\label{fig:missing_data}
\end{figure}

{\it Simulation results }
Figure~\ref{fig:missing_data} summarizes the empirical squared error (SE) of all estimators over $R=500$ Monte Carlo replications for each sample size. As expected, under the correctly specified MAR setting, all estimators exhibit decreasing SE as the sample size $N$ increases, reflecting consistency and asymptotic efficiency in this well-behaved regime.
When both the outcome regression and the propensity score models are correctly specified, classical estimators such as PL, IPW, AIPW, and DML perform similarly, with no single method uniformly dominating the others across sample sizes.
In this setting, the proposed W-V and data-splitting-based W-estimator variants (W-S) in Algorithm~\ref{algorithm-data-splitting-missing} achieve SE levels that are comparable to those of the classical and doubly robust estimators, confirming that the proposed approach does not incur an efficiency loss in canonical MAR scenarios where standard methods are known to be optimal.
Notably, the W-S and W-V estimators exhibit stable finite-sample behavior across all sample sizes considered, with variability comparable to that of AIPW and DML.

Taken together with the results from the covariate shift experiments in the previous subsection, these findings demonstrate that the proposed estimators adapt seamlessly across regimes: they match the performance of classical estimators in idealized MAR settings while offering improved robustness and favorable performance under model misspecification.
This combination of efficiency preservation and robustness highlights the practical appeal of the proposed approach for missing data problems encountered in realistic settings.

\section{Real data analysis}
\label{sec-real}
We further demonstrate the finite-sample performance of the proposed estimator using a real dataset from OpenML~\citep{feurer2021openml}, \emph{rainfall\_bangladesh} (ID 41263). 
The dataset consists of historical rainfall measurements collected by the Bangladesh Meteorological Department across multiple weather stations, together with auxiliary information such as station location and year. 
From a missing data perspective, this dataset is particularly relevant because rainfall measurements are often incomplete in practice due to monitoring limitations, reporting delays, or station-specific operational issues, while auxiliary covariates remain fully observed. 
This setting naturally aligns with a MAR framework, in which the probability of observing rainfall may depend on observed covariates but not on the unobserved response conditional on those covariates.

The inferential target is the marginal mean of the response variable, namely the amount of rainfall. 
Since the true population mean is unknown in practice, we adopt a semi-real experimental design commonly used in the missing data literature. 
Specifically, we treat the empirical mean computed from the full dataset as the ground-truth parameter value. 
To mimic realistic sampling variability, we repeatedly draw simple random samples of size $N=6000$ from the full dataset. 
For each sampled dataset, we generate missing responses under a MAR mechanism by specifying a propensity score that depends on the observed year variable. 
In particular, the propensity score is given by $\pi(x) = 1 - \text{expit}(x^2-1)$ where $x$ denotes the (appropriately normalized) year covariate, and response indicators $D_i$ are generated independently according $\pi(x_i)$.
The entire sampling and missingness generation procedure is repeated 100 times. 

For all methods, we use linear models for both the outcome regression and the propensity score estimation, as linear regression has been widely studied and applied to this dataset according to its OpenML description. 
For each repetition, we apply the proposed method and several competing estimators to estimate the response mean, and we evaluate their performance using the squared error relative to the ground-truth value. 
This design enables a systematic comparison of robustness and accuracy across methods under realistic covariate distributions and empirically calibrated missingness mechanisms. 
Let $\hat\theta_r$ denote the estimate of the true parameter $\theta^*$ in the $r$th repetition. Table~\ref{tab:real_data} reports the mean and standard deviation of the SE $(\hat\theta_r - \theta^*)^2$ across 100 repetitions.

\begin{table}[!ht]
\centering
\caption{Averages and standard deviations of the SEs across 100 repetitions for different estimators on the \emph{rainfall\_bangladesh} dataset. Values are scaled by $10^4$ relative to the raw data.}
\label{tab:real_data}
\begin{tabular}{llllllll}
\toprule
Estimators & Na\"ive & PL & IPW & AIPW & DML & \textbf{W-V} & W-S \\
\midrule
SE (average) &  3.9771 & 3.0763 & 3.1657 & 3.0711 & 3.1275 & \textbf{2.6581} & 3.0755\\
SE (standard deviation)  & 4.4454 & 3.8721 & 3.9985 & 3.8523 & 3.9459 & \textbf{3.8300} &  3.8714\\
\bottomrule
\end{tabular}
\end{table}
The results indicate that the proposed W-V estimator achieves the lowest mean squared error among all methods considered and the SE has a smaller variation compared to other approaches.
The W-S estimator exhibits slightly inferior performance but remains competitive with existing approaches. 
All methods substantially outperform the Na\"ive estimator, which relies solely on complete cases for estimating the response mean.

\section{Discussion} 
\label{sec-discussion}
We propose a novel minimum Wasserstein distance estimator (W-estimator) for parameter estimation under covariate shift and missing data settings. We establish that the proposed W-estimator is asymptotically normal and achieves asymptotic super-efficiency under certain conditions, despite requiring no parametric model assumptions. This super-efficiency is likely attributable to its irregular asymptotic behavior, i.e. it is not asymptotically linear. 
We demonstrate that the proposed W-estimator is equivalent to the 1-NN estimator; hence, our findings imply that the 1-NN estimator is also asymptotically normal and may exhibit asymptotic super-efficiency. Moreover, the W-estimator provides an optimal transport interpretation of the 1-NN estimator. In contrast to the 1-NN estimator, our W-estimator can be flexibly enhanced by incorporating pre-trained statistical or machine learning regression models. We prove that the enhanced estimator generally retains the same limiting distribution as the original W-estimator, regardless of the regression model used. 

Our proposed minimum Wasserstein distance estimation method naturally leads to a W-estimator   $F_1(\bx, y; \hat \bp)$ for the target population distribution  $F_1(\bx, y)$. Building on this W-estimator, one can readily construct estimators for any functional of $F_1(\bx, y)$. To study the large-sample properties of such estimators, a practical strategy is to first characterize the limiting behavior of the stochastic process
$ 
\left\{ \sqrt{m} \left( F_1(\bx, y; \hat \bp) - F_1(\bx, y) \right) : \bx \in \mathcal{X},\ y \in \mathcal{Y} \right\}
$
and then apply the continuous mapping theorem. However, this first step poses a nontrivial challenge, as the asymptotic behavior of $F_1(\bx, y; \hat \bp)$ may be irregular, similar to that observed for $\hat{\theta}$.  This topic remains open for future investigation.

\section*{Acknowledgements}
This work is supported in part by   the National Natural Science Foundation of China (12571283 and 12301391), the Fundamental and Interdisciplinary Disciplines Breakthrough Plan of the Ministry of Education of China (JYB2025XDXM904), and the National Key R\&D Program of China Grant 2024YFA1015800. 
Zhang and Liu are co-corresponding authors.

\bibliographystyle{natbib}
\bibliography{MWLE}

\end{document}